\documentclass[journal]{IEEEtran}

\usepackage{breqn}
\usepackage{amsthm,amsmath,amssymb}
\usepackage{caption}
\usepackage{graphicx}
\usepackage{booktabs}
\usepackage{tabularray}
\usepackage{etoolbox}
\usepackage{stfloats}
\usepackage[font=small]{caption}
\usepackage{cite}
\usepackage{booktabs}
\usepackage[labelformat=simple]{subcaption}
\usepackage{soul}
\usepackage{tabularx}
\usepackage{textcomp}
\usepackage{mathtools}
\usepackage[ruled,vlined,linesnumbered]{algorithm2e}
\usepackage{color}
\usepackage[dvipsnames]{xcolor}
\usepackage[normalem]{ulem}
\newcommand\soutpars[1]{\let\helpcmd\sout\parhelp#1\par\relax\relax}
\long\def\parhelp#1\par#2\relax{%
  \helpcmd{#1}\ifx\relax#2\else\par\parhelp#2\relax\fi%
}

\usepackage{array}
\newcommand{\PreserveBackslash}[1]{\let\temp=\\#1\let\\=\temp}
\newcolumntype{C}[1]{>{\PreserveBackslash\centering}m{#1}}
\newcolumntype{R}[1]{>{\PreserveBackslash\raggedleft}m{#1}}
\newcolumntype{L}[1]{>{\PreserveBackslash\raggedright}m{#1}}

\makeatletter 
\pretocmd\@bibitem{\color{black}\csname keycolor#1\endcsname}{}{\fail}
\newcommand\citecolor[1]{\@namedef{keycolor#1}{\color{blue}}}
\makeatother

\UseTblrLibrary{booktabs,siunitx}

\DeclareSIUnit{\belmilliwatt}{Bm}
\DeclareSIUnit{\dBm}{\deci\belmilliwatt}
\DeclareSIUnit{\rad}{rad}

\ifCLASSINFOpdf

\else

\fi
\newtheorem{assump}{Assumption}

\newtheorem{remark}{Remark}

\SetKwInOut{KwIni}{Initialization}

\allowdisplaybreaks

\begin{document}
\bstctlcite{IEEEexample:BSTcontrol}
\title{Array Partitioning Based Near-Field Attitude and Location Estimation}
%
%

\author{Mingchen~Zhang,~\IEEEmembership{Student~Member,~IEEE,}
        Xiaojun~Yuan,~\IEEEmembership{Senior~Member,~IEEE,} 
        Boyu~Teng,~\IEEEmembership{Student~Member,~IEEE,}
        and Li Wang,~\IEEEmembership{Senior~Member,~IEEE} 
\thanks{M. Zhang, X. Yuan, B. Teng, and L. Wang are with the National Key Laboratory of Wireless Communications, University of Electronic Science and Technology of China, Chengdu 611731, China (e-mail: mczhang@std.uestc.edu.cn; xjyuan@uestc.edu.cn; byteng@std.uestc.edu.cn; powerking@live.co.uk). This paper was presented in part at the 16th International Conference on Wireless Communications and Signal Processing (WCSP 2024), Hefei, China \cite{zhang_2024_WCSP}.}
}


\maketitle

\begin{abstract}
    This paper studies a {\color{black}passive source localization} system, where a single base station (BS) is employed to estimate the positions and attitudes of multiple mobile stations (MSs). The BS and the MSs are equipped with uniform rectangular arrays, and the MSs are located in the near-field region of the BS array. To avoid the difficulty of tackling the problem directly based on the near-field signal model, we establish a subarray-wise far-field received signal model. In this model, the entire BS array is divided into multiple subarrays to ensure that each MS is in the far-field region of each BS subarray. By exploiting the angles of arrival (AoAs) of an MS antenna at different BS subarrays, we formulate the attitude and location estimation problem under the Bayesian inference framework. Based on the factor graph representation of the probabilistic problem model, a message passing algorithm named array partitioning based pose and location estimation (APPLE) is developed to solve this problem. An estimation-error lower bound is obtained as a performance benchmark of the proposed algorithm. Numerical results demonstrate that the proposed APPLE algorithm outperforms other baseline methods in the accuracy of position and attitude estimation.
\end{abstract}


\IEEEpeerreviewmaketitle

\section{Introduction}

{\color{black}
Over the past few decades, the multiple-input multiple-output (MIMO) technology has been a key driver in enhancing the performance of wireless communication systems. 
By deploying antenna arrays at the transceivers, and using multiple antennas to simultaneously transmit and receive signals, the MIMO technology significantly enhances spectral efficiency and data transmission rates. 
Beyond communication, wireless positioning has also benefited from the MIMO technology. The joint operation of multiple antennas in an array provides higher spatial resolution, enabling more precise estimation of parameters such as angle of arrival (AoA) and time of arrival (ToA). 
Currently, significant research is focused on integrating advanced MIMO technologies, such as mmWave/Terahertz MIMO, intelligent reflecting surface (IRS), and extremely large-scale MIMO (XL-MIMO), to further enhance wireless positioning services \cite{garciaDirectLocalizationMassive2017,savicFingerprintingBasedPositioningDistributed2015,tengBayesianUserLocalization2022,heLargeIntelligentSurface2020a,rinchiCompressiveNearFieldLocalization2022a}.
}

Recently, wireless localization in \textit{near-field} scenarios has attracted intensive interest, driven by the further expansion of the utilized frequency bands and antenna array scale in 6G networks{\color{black}\cite{rinchiCompressiveNearFieldLocalization2022a,dardariNLOSNearFieldLocalization2022b,wangNearFieldIntegratedSensing2023,huaNearField3DLocalization2024,panRISAidedNearFieldLocalization2023}}. For an antenna array, the surrounding space can be generally divided into the far-field region (FFR) and the near-field region\footnotemark\footnotetext{In this paper, the term ``near-field'' specifically refers to the Fresnel region (also known as the radiative near-field). The reactive near-field region is neglected for its very limited range (typically on the order of wavelengths).}(NFR), determined by whether the spherical wavefront radiated by a point source located in the region can be approximated as a plane upon reaching the antenna array. A widely-accepted distance boundary for dividing the NFR and the FFR is the so-called Rayleigh distance \cite{selvanFraunhoferFresnelDistances2017}. 6G wireless networks are expected to operate at mmWave or even THz bands, and to employ extremely large-scale antenna arrays (ELAAs) comprising up to ten thousands of antennas \cite{wangTutorialExtremelyLargeScale2024}. Due to the severe path loss in mmWave and THz bands, the coverage radius of 6G cellular cells is limited to a few tens of meters \cite{saadVision6GWireless2020a}. However, the Rayleigh distance of a 6G base station (BS) employing an ELAA can extend to hundreds of meters or more \cite{cuiNearFieldMIMOCommunications2023}. 
This means that user terminals in a cell may always fall in the NFR of the BS.
Directly applying traditional far-field localization methods to near-field scenarios can lead to significant performance degradation due to model mismatch.
To address this issue, various near-field localization techniques have been developed in recent years, including those based on angle of arrival (AoA){\color{black}\cite{liangPassiveLocalizationMixed2010a,zhengScalableNearFieldLocalization2023}}, time of arrival (ToA) \cite{dardariNLOSNearFieldLocalization2022b}, time difference of arrival (TDoA){\color{black}\cite{wangConvexRelaxationMethods2019,sunSolutionAnalysisTDOA2019}}, and others.
Existing wireless localization methods typically focus on point-source localization, where the MS is regarded as a mass point. However, in many localization applications, especially in near-field scenarios, the MS for positioning needs to be regarded as a rigid body with unignorable physical size and shape. Furthermore, localizing the MS involves estimating both its position and attitude. For instance, in the context of indoor robot navigation, knowledge of both the robot's position and attitude is necessary for achieving precise steering control. This extended localization problem is referred to as position and attitude estimation (PAE), also known as position and orientation estimation \cite{shahmansooriPositionOrientationEstimation2018}, position and pose estimation \cite{marchandPoseEstimationAugmented2016a}, rigid body localization \cite{wangInvestigationSolutionAngle2020}, and six-dimensional positioning \cite{bjornsonMassiveMIMOReality2019a}.

Various approaches have been proposed to address the PAE problem, as described in \cite{talvitieHighAccuracyJointPosition2019,liJointLocalizationOrientation2022a,mendrzikHarnessingNLOSComponents2019a,zhouDoABasedRigidBody2019}. Among these approaches, methods based on AoA estimation, such as those proposed in \cite{zhouDoABasedRigidBody2019,wangInvestigationSolutionAngle2020,wangBiasReducedSemidefinite2021}, stand out for their high localization accuracy without requiring wideband signaling or strict synchronization between the BS and the MS. In these studies, multiple sensors are attached to the MS and communicate with the BS, with their relative positions predetermined and known. The PAE problem is formulated as the concatenation of two sub-problems. The first sub-problem is to estimate the AoAs of the sensors at the BS. The second is to estimate MS's position and attitude based on the AoA estimates and the known topological knowledge of the sensor array. The two sub-problems are addressed sequentially by following a two-stage mechanism.

The existing AoA-based PAE methods however have their limitations. Firstly, as the AoA estimation sub-problem and the AoA-based PAE sub-problem are tackled separately, the two-stage approach fails to fully utilize the prior topological information of the sensors, i.e., the topological information of the sensor array is not used to enhance the AoA estimation. Therefore, the estimation performance of the two-stage approach is fundamentally compromised. 
Another issue arises in the interested near-field scenarios. Most existing methods for solving the AoA estimation sub-problem are based on the far-field assumption, i.e., the MS is in the FFR of the BS array, and the entire BS array sees a common AoA from a sensor at its reference point. However, when the MS is in the NFR of the BS array, due to the spherical wave characteristic, the AoA of the signals from a sensor varies among the BS array, making it challenging to determine the AoA at a specific reference point. Directly applying far-field AoA estimation methods to near-field scenarios can result in significant performance degradation, thereby limiting the PAE accuracy of the two-stage methods.
Given these considerations, in the near-field scenarios, it is desirable to formulate a unified PAE problem rather than treating it as separate sub-problems. A viable solution that can fully exploit the topological knowledge of the sensors and the AoA variations at different elements of the BS array is also required.

In this paper, we present a problem formulation for the near-field PAE of multiple MSs and propose a novel algorithm to solve the problem. The proposed method borrows the idea of array partitioning in \cite{zhengScalableNearFieldLocalization2023}.
Specifically, we consider a scenario where a single BS is used for the PAE of multiple MSs. Both the BS and each MS are equipped with a uniform rectangular array (URA). During the localization phase, certain antennas on each MS array are activated to transmit signals. The BS aims to estimate the positions and attitudes of the MSs based on the received signals and the prior topological knowledge of the activated MS antennas. Given the inherent difficulty of this task under the original near-field signal model, we partition the BS array into multiple subarrays and introduce a \textit{subarray-wise far-field (SWFF) signal model}, where the activated MS antennas are in the FFR of each BS subarray. Based on this signal model, we formulate a probabilistic PAE problem by exploiting the AoA difference of received signals at BS subarrays and the prior topological knowledge of the activated MS antennas. Message passing is performed based on the probability model of the problem. Laplace approximation is introduced in the algorithm design to handle messages difficult to compute. The proposed algorithm is named as array partitioning based pose and location estimation (APPLE). For evaluating the performance of the APPLE algorithm, we derive an estimation-error lower bound of the considered PAE problem. 
Numerical results demonstrate that APPLE exhibits superior estimation accuracy and can closely approach the error lower bound. The contributions of this paper are summarized as follows:
\begin{itemize}
    \item By adopting the array partitioning methodology, we formulate a probabilistic PAE problem based on the SWFF signal model. 
    \item By leveraging a factor graph representation of the SWFF signal model, we develop the APPLE algorithm for solving the considered PAE problem.
    \item An estimation-error lower bound for the considered PAE problem is obtained. It serves as a performance benchmark of the proposed APPLE algorithm.
    \item We show that the proposed APPLE algorithm significantly outperforms other baseline methods in estimation accuracy under different parameter settings. The PAE performance by APPLE can closely approach the derived estimation-error lower bound.
\end{itemize}

{\color{black}
\subsection{Related Work}
Traditionally, position and attitude estimation are treated as separate tasks, with specialized techniques developed for attitude estimation. These include methods based on inertial sensors \cite{esfahaniOriNetRobust3D2020}, stereo vision \cite{zhangRobustMethodMeasuring2021}, and magnetometers \cite{zhangMicromagnetometerCalibrationAccurate2015}. However, these approaches require additional sensors and are sensitive to noise and environmental interference. For instance, inertial sensors are vulnerable to temperature fluctuations and mechanical vibrations, which can lead to cumulative errors (drift), degrading the estimation accuracy over time \cite{kokUsingInertialSensors2018}. Similarly, magnetometer-based methods are susceptible to magnetic field disturbances, while vision-based techniques are sensitive to lighting conditions. In contrast, the proposed method is based on wireless signals, which are less influenced by these environmental factors and can be effectively integrated with communication devices, thus eliminating the need for extra sensors. 
As an efficient Bayesian inference technique, the message passing algorithm, also known as belief propagation or the sum-product algorithm, has been applied to positioning problems in various scenarios.
For instance, 
Ref. \cite{meyerDistributedLocalizationTracking2016} proposes a message passing based approach for distributed sequential localization and tracking in mobile networks, integrating cooperative self-localization and distributed tracking for both mobile agents and noncooperative entities.
Ref. \cite{yuanCooperativeJointLocalization2016} introduces a variational message passing algorithm for joint localization and time synchronization using ToA measurements. Ref. \cite{tengBayesianUserLocalization2022} addresses user tracking in IRS-aided MIMO systems, employing a message passing algorithm to estimate MSs' positions in real-time while optimizing beamforming strategies at the BS and IRS. 
In this work, we design a message passing algorithm to solve a near-field PAE problem. The proposed algorithm can effectively integrate prior knowledge and the AoA estimates from the BS subarrays, enabling the joint estimation of MSs' positions and attitudes.
}

The remainder of this paper is organized as follows. In Section II, we introduce the considered system and the near-field received signal model. In Section III, we present the received signal model based on the SWFF assumption and then formulate the PAE problem. In Section IV, we develop the proposed APPLE algorithm under the message passing framework. In Section V, we derive an error lower bound for the considered PAE problem. Numerical results are provided in Section VI, and conclusions are drawn in Section VII.

\textit{Notation:} We use a bold symbol lowercase letter and bold symbol capital letter to denote a vector and a matrix, respectively. The transpose, conjugate transpose, and inverse of a matrix are denoted by $(\cdot)^{\mathsf{T}}$, $(\cdot)^{\mathsf{H}}$, and $(\cdot)^{-1}$, respectively; $||\cdot||$ denotes the $\ell^2$ norm; {\color{black} $\circ $} denotes the cross product operator; {\color{black} $\delta(\cdot)$ denotes the Dirac delta function;} $\mathbb{E}_p[\boldsymbol{x}]$ denotes the expectation of $\boldsymbol{x}$ over distribution $p$; $\mathcal{I}_{N}$ denotes the index set $\{1,\ldots,N\}$, where $N$ is a positive integer.

\section{System Model}

\subsection{System Description}



\begin{figure}[t]
    \centering
    \includegraphics[width=.88\linewidth]{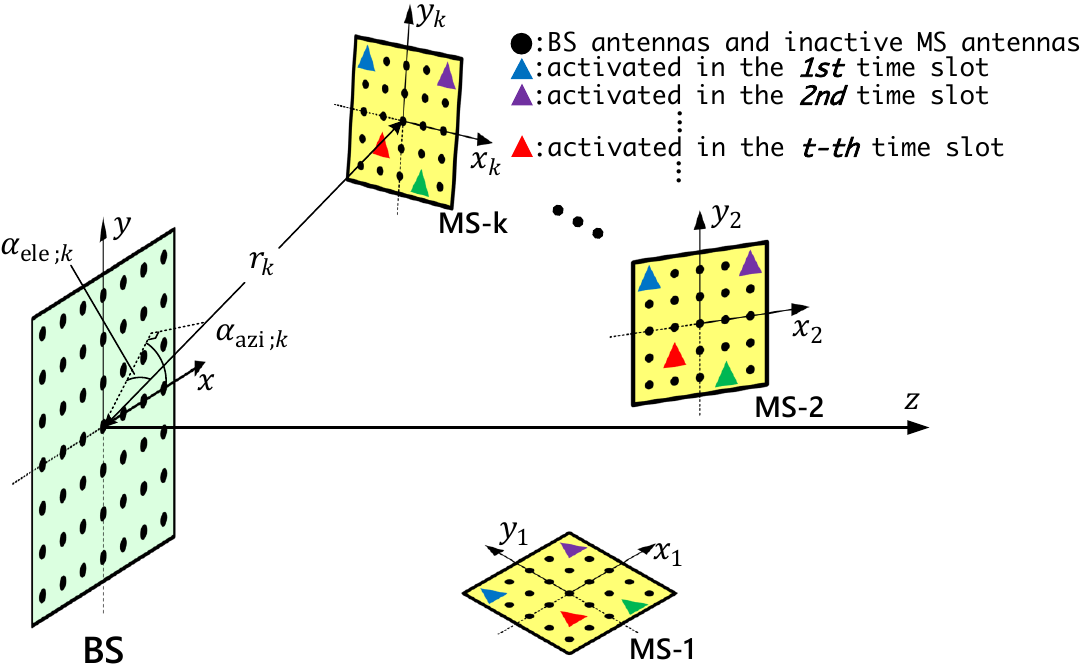}
    \caption{The considered near-field multiuser localization system.}
    \label{scene}
\end{figure}

We consider a multiuser MIMO system consisting of one BS and $K$ MSs, as illustrated in Fig. \ref{scene}. The BS and each MS are equipped with URAs containing $N_{\mathsf{B}}$ and $N_{\mathsf{M}}$ antenna elements, respectively. The antenna spacings at the BS and the MS array are set to half of the carrier wavelength $\lambda$. The BS array is an ELAA with a relatively large $N_{\mathsf{B}}$. Establish a \textit{global} coordinates $O$-$x$-$y$-$z$ with the origin $O$ located at the center of the BS array and the $x$- and $y$- axes aligned with the edges of the BS array. The $z$-axis is generated following the right-hand rule with respect to the $\mathsf{x}$- and $\mathsf{y}$- axes. 
Denote by $\boldsymbol{e}_{\mathsf{x}}$, $\boldsymbol{e}_{\mathsf{y}}$, and $\boldsymbol{e}_{\mathsf{z}}$ the unit vectors of the $\mathsf{x}$-, $\mathsf{y}$-, and $\mathsf{z}$-axes, respectively. 
The $N_{\mathsf{B}}$ antennas at the BS array are evenly arranged on an $N_{\mathsf{B};\mathsf{x}}\times N_{\mathsf{B};\mathsf{y}}$ grid, where $N_{\mathsf{B}}=N_{\mathsf{B};\mathsf{x}}N_{\mathsf{B};\mathsf{y}}$. The size of the BS array is $S_{\mathsf{B};\mathsf{x}}\times S_{\mathsf{B};\mathsf{y}}$. The global coordinate of the BS antenna located at the $u$-th position along the $x$-axis and the $v$-th position along the $y$-axis (refer to as the $(u,v)$-th BS antenna) is given by 
\begin{equation}
    \mathbf{p}_{\mathsf{B};(u,v)} \! = \! \left[ \!\left( \!u-\frac{N_{\mathsf{B};\mathsf{x}}+1}{2} \! \right)\frac{\lambda}{2},\left( \! v-\frac{N_{\mathsf{B};\mathsf{y}}+1}{2} \!\right)\frac{\lambda}{2},0 \right]^{\mathsf{T}} \!\!,
\end{equation}
$u\in \mathcal{I}_{N_{\mathsf{B};\mathsf{x}}}$, $v\in\mathcal{I}_{N_{\mathsf{B};\mathsf{y}}}$.

For the MS arrays, denote by $O_k$ the center of the $k$-th MS array, and by $\mathbf{p}_{\mathsf{M};k}$ the position of $O_k$ in the global coordinates, $k\in \mathcal{I}_K$. 
For the $k$-th MS array, a \textit{local} coordinates $O_k$-$x_k$-$y_k$ can be established with the origin located at $O_k$, and the $x_k$- and $y_k$- axes aligned with the array edges. The $N_{\mathsf{M}}$ antennas at each $k$-th MS array are evenly arranged on an $N_{\mathsf{M};\mathsf{x}} \times N_{\mathsf{M};\mathsf{y}}$ grid.
The local coordinates of the antenna located at the $q$-th position in the $x_k$-direction and the $s$-th position in the $y_k$-direction of the $k$-th MS array (refer to as the $(k,q,s)$-th MS antenna) can be expressed as
\begin{equation}\label{qMkqs}
    \mathbf{q}_{\mathsf{M};(k,q,s)} \! = \! \left[\!\left(q-\frac{N_{\mathsf{M};\mathsf{x}}+1}{2}\right) \! \frac{\lambda}{2},\left(s-\frac{N_{\mathsf{M};\mathsf{y}}+1}{2}\right) \! \frac{\lambda}{2}  \right]^{\mathsf{T}} \!\!,
\end{equation}
$\forall k\in \mathcal{I}_{K}$, $q\in \mathcal{I}_{N_{\mathsf{M};\mathsf{x}}}$, $s\in \mathcal{I}_{N_{\mathsf{M};\mathsf{y}}}$.
The local coordinates of each MS array can be considered as obtained through a combination of rotation and translation from the global coordinates. Specifically, denote by $\boldsymbol{e}_{x_k}$ and $\boldsymbol{e}_{y_k}$ the unit vectors of the $x_{k}$- and $y_k$- axes, respectively. $\boldsymbol{e}_{x_k}$ and $\boldsymbol{e}_{y_k}$ can be represented as 
\begin{align}\label{eRTransform}
    \left[\boldsymbol{e}_{x_k},\boldsymbol{e}_{y_k}\right] &= \mathbf{R}(\boldsymbol{\theta}_k), 
\end{align}
where $\boldsymbol{\theta}_k$ is a vector given by $\boldsymbol{\theta}_k \triangleq \left[\theta_{\mathsf{x};k},\theta_{\mathsf{y};k},\theta_{\mathsf{z};k} \right]^{\mathsf{T}}$, with $\theta_{\mathsf{x};k}\in [-\pi,\pi)$ (roll), $\theta_{\mathsf{y};k}\in\left[-\frac{\pi}{2},\frac{\pi}{2}\right]$ (pitch), and $\theta_{\mathsf{z};k}\in [-\pi,\pi)$ (yaw) being the rotation angles of the local $O_k$-$x_k$-$y_k$ coordinates over the global $O$-$x$-$y$ coordinates.
$\mathbf{R}(\boldsymbol{\theta}_k)$ is the corresponding rotation matrix given by 
\begin{align}
    &\mathbf{R}(\boldsymbol{\theta}_k)= \label{Rdef} \\
    &\begin{bmatrix}
        \cos \theta_{\mathsf{z};k} \cos \theta_{\mathsf{y};k} & \cos \theta_{\mathsf{z};k} \sin \theta_{\mathsf{y};k} \sin \theta_{\mathsf{x};k} - \sin \theta_{\mathsf{z};k} \cos \theta_{\mathsf{x};k}  \\
        \sin \theta_{\mathsf{z};k} \cos \theta_{\mathsf{y};k} & \sin \theta_{\mathsf{z};k} \sin \theta_{\mathsf{y};k} \sin \theta_{\mathsf{x};k} + \cos \theta_{\mathsf{z};k} \cos \theta_{\mathsf{x};k}  \\
        -\sin \theta_{\mathsf{y};k} & \cos \theta_{\mathsf{y};k} \sin \theta_{\mathsf{x};k}
    \end{bmatrix}. \notag
\end{align}
Then, based on \eqref{qMkqs} and \eqref{eRTransform}, the global coordinates of the $(k,q,s)$-th MS antenna is given by 
\begin{equation}\label{pMkqs}
    \mathbf{p}_{\mathsf{M};(k,q,s)} = \mathbf{p}_{\mathsf{M};k} + \mathbf{R}(\boldsymbol{\theta}_k) \mathbf{q}_{\mathsf{M};(k,q,s)},
\end{equation}
$k\in \mathcal{I}_{K}$, $q\in \mathcal{I}_{N_{\mathsf{M};\mathsf{x}}}$, $s\in \mathcal{I}_{N_{\mathsf{M};\mathsf{y}}}$.

\subsection{Channel Model}
Since the ELAAs are typically considered for operation in the millimeter-wave or even terahertz frequency bands, where the channel strength disparity between line-of-sight (LoS) and non-line-of-sight (NLoS) links can be over 20 dB, we solely consider the LoS channel modelling between the BS and the MSs{\color{black}\cite{liuLineofSightSpatialModulation2016}}. Besides, we assume that all antennas of $K$ MS arrays are out of the reactive near field of the BS array, i.e., the distances between the MS antennas and the origin $O$ satisfying
\begin{equation}\label{NFR}
    r_{k,q,s} \triangleq \|\mathbf{p}_{\mathsf{M};(k,q,s)}\| \geq  D_{\mathsf{F}},~ \forall k\in \mathcal{I}_K, ~\forall q \in \mathcal{I}_{N_{\mathsf{M};\mathsf{x}}}, ~ \forall s\in \mathcal{I}_{N_{\mathsf{M};\mathsf{y}}},
\end{equation}
where $D_{\mathsf{F}} \triangleq \sqrt[3]{\frac{\tilde{S}_{\mathsf{B}}^4}{8\lambda}}$ is known as the Fresnel distance of the BS array, and $\tilde{S}_{\mathsf{B}}$ is the largest dimension of the BS array \cite{selvanFraunhoferFresnelDistances2017}. For practical antenna array sizes, the Fresnel distance of an antenna array is generally short. For example, taking $\tilde{S}_{\mathsf{B}}=0.5$m and $\lambda=0.004$m (with a carrier frequency $f=75$GHz in the millimeter wave band) yields a Fresnel distance of $D_{\mathsf{F}}=1.25$m. Eqn. \eqref{NFR} means that the MS antennas may reside in the NFR of the BS array, where the curvatures of the spherical wavefronts radiated by the MS antennas must be taken into account at the BS array.  
A widely-accepted boundary between the NFR and the FFR of an antenna array is the Rayleigh distance \cite{selvanFraunhoferFresnelDistances2017}, given by
\begin{equation}
    D_{\mathsf{R}}\triangleq \frac{2\tilde{S}_{\mathsf{B}}^2}{\lambda}.
\end{equation}
The MS antennas falling within the range $D_{\mathsf{F}}\leq r_{k,q,s} \leq D_{\mathsf{R}}$ are all in the NFR of the BS array. The NFR of a BS array can cover a large area. Still taking $\tilde{S}_{\mathsf{B}}=0.5$m and $\lambda=0.004$m as an example, the Rayleigh distance extends up to $D_{\mathsf{R}}=125$m.


{\color{black}
We consider that the BS and MS antennas are modelled as isotropic with a uniform gain in all directions. Then, by following \cite{luTutorialNearFieldXLMIMO2024}, the channel coefficient between the $(u,v)$-th BS antenna and the $(k,q,s)$-th MS antenna is modelled as
\begin{equation}\label{hcoeff}
    h_{(u,v),(k,q,s)} = \beta_k \rho_{(u,v),(k,q,s)} e^{-\frac{j2\pi r_{(u,v),(k,q,s)}}{\lambda}},
\end{equation}
where $\lambda$ is the carrier wavelength; $\beta_k$ is the antenna gain between the BS array and the $k$-th MS array; $r_{(u,v),(k,q,s)} = \|\mathbf{p}_{\mathsf{M};(k,q,s)} - \mathbf{p}_{\mathsf{B};(u,v)}\|$ denotes the distance between the $(u,v)$-th BS antenna and the $(k,q,s)$-th MS antenna; $\rho_{(u,v),(k,q,s)} = \frac{\lambda}{4\pi r_{(u,v),(k,q,s)}}$ represents the free space path loss, $u\in \mathcal{I}_{N_{\mathsf{B};\mathsf{x}}}$, $v\in\mathcal{I}_{N_{\mathsf{B};\mathsf{y}}}$, $k\in \mathcal{I}_{K}$, $q\in \mathcal{I}_{N_{\mathsf{M};\mathsf{x}}}$, $s\in \mathcal{I}_{N_{\mathsf{M};\mathsf{y}}}$.}
The channel model given in \eqref{hcoeff} takes into account the exact link distances to describe the phase shifts of the channel coefficients, instead of adopting the plane-wave assumption commonly used in far-field channel models.


\subsection{Signal Model}\label{secSigModel}

At the localization phase, the MSs transmit signals to the BS for position and attitude estimation. 
We consider a transmission strategy in which each MS array activates part of its antennas to transmit signals, and all the MS arrays share a common index set of the activated antennas. 
Specifically, define the activated MS antenna index set as $\mathcal{T} \subset \{(q,s)|q\in \mathcal{I}_{N_{\mathsf{M};\mathsf{x}}},s\in \mathcal{I}_{N_{\mathsf{M};\mathsf{y}}} \}$, which is a subset of the overall MS antenna index set.
The cardinality of $\mathcal{T}$ is denoted by $T$. 
For ensuring the distinguishability at the BS, we consider that the activated antennas of an MS array use orthogonal time or frequency resources to transmit signals. Meanwhile, activated antennas with identical index across different MS arrays share time and frequency resources for signal transmission. Such a transmission strategy is illustrated in Fig. \ref{scene}, where black circles denote BS antennas and inactive MS antennas, and triangles denote activated MS antennas. Activated antennas on a MS array are distinguished by color, and the activated antennas with the same index across different MS arrays are assigned the same color.
In this paper, we henceforth adopt the time-division transmission strategy, where a total of $T$ time slots are used for signal transmission. 
Denote by $\boldsymbol{l}_t \triangleq [q(t),s(t)]^{\mathsf{T}}$ the index vector of the antennas activated in the $t$-th time slot, and define an index matrix $\boldsymbol{L} \triangleq \left[\boldsymbol{l}_1, \ldots, \boldsymbol{l}_T \right] \in \mathbb{R}^{2\times T}$.
$\boldsymbol{L}$ represents the topological information of the activated antennas at the MSs, and is referred to as the \textit{transmit pattern}. The value of $\boldsymbol{L}$ is predetermined between the MSs and the BS before the signal transmission.
Denote by $x_{k,t}$ the symbol transmitted by the $(k,q(t),s(t))$-th MS antenna, and by 
\begin{multline}
    \boldsymbol{h}_{(k,q(t),s(t))}
     \triangleq \\
      \left[h_{(1,1),(k,q(t),s(t))},\ldots,h_{(N_{\mathsf{B};\mathsf{x}},N_{\mathsf{B};\mathsf{y}}),(k,q(t),s(t))}\right]^{\mathsf{T}}\in \mathbb{C}^{N_{\mathsf{B}}}
\end{multline}
the channel between the BS array and the $(k,q(t),s(t))$-th MS antenna. For notational brevity, we henceforth abbreviate the activated MS antenna index $(k,q(t),s(t))$ as $(k,t)$. The MSs are assumed to be synchronized through mechanisms such as timing advance \cite{3gpp.38.213}.
In the $t$-th time slot, at the $k$-th MS, the symbol $x_{k,t}$ is pulse shaped and transmitted by the radio frequency (RF) chain. Upon reception at the BS, the baseband received signal model can be expressed as 
\begin{equation}\label{SigModel}
    \boldsymbol{y}_t = \sum_{k\in\mathcal{I}_K} \boldsymbol{h}_{(k,t)} x_{k,t} + \boldsymbol{w}_t,
\end{equation}
where $\boldsymbol{w}_t \sim \mathcal{CN}(\boldsymbol{0},\sigma_w^2 \boldsymbol{I})$ is the circularly symmetric complex Gaussian (CSCG) noise with $\sigma_w^2$ being the noise variance. 
The received signals in $T$ time slots can be stacked into a matrix as
\begin{equation}\label{Y}
    \boldsymbol{Y}=[\boldsymbol{y}_{1},\ldots,\boldsymbol{y}_{T}]\in \mathbb{C}^{N_{\mathsf{B}}\times T}.
\end{equation}
Eqn. \eqref{Y} is referred to as the \textit{near-field} received signal model. 
We aim to estimate the position and the attitude of each MS, i.e., to estimate the parameter set $\mathcal{P}_{\boldsymbol{e}}\triangleq \{\mathbf{p}_{\mathsf{M};k}, \boldsymbol{e}_{x_k}, \boldsymbol{e}_{y_k} | k\in\mathcal{I}_{K}\}$, or equivalently, to estimate $\mathcal{P}_{\boldsymbol{\theta}}\triangleq \{\mathbf{p}_{\mathsf{M};k}, \boldsymbol{\theta}_{k} | k\in\mathcal{I}_{K}\}$
based on the received signal $\boldsymbol{Y}$ and the transmit pattern $\boldsymbol{L}$. A straightforward approach to the PAE problem is to compute the maximum likelihood (ML) estimate based on the near-field received signal model \eqref{Y}. 
This ML problem can be formulated as 
\begin{equation}\label{MLproblem}
    \max_{\mathcal{P}_{\boldsymbol{\theta}}}~ p(\boldsymbol{Y};\mathcal{P}_{\boldsymbol{\theta}}).
\end{equation}
Problem \eqref{MLproblem} is difficult to solve. A simplified problem for the single-MS and single-snapshot case has been previously studied in \cite{zhengScalableNearFieldLocalization2023}, where the MS is reduced to a mass point and the position of the MS, denoted by $\mathbf{p}_{\mathsf{M};1}$, is to be estimated at the BS. It has been shown in \cite{zhengScalableNearFieldLocalization2023} that the corresponding likelihood $p(\boldsymbol{Y};\mathbf{p}_{\mathsf{M};1})$ is highly multi-modal, and the peak becomes sharper as the array size increases. It is thus difficult to determine the evasive peak of $p(\boldsymbol{Y};\mathbf{p}_{\mathsf{M};1})$.
The extension to joint estimation of the positions and attitudes of multiple MSs in \eqref{MLproblem} is clearly more challenging. In this paper, we aim to develop a method for accurate and efficient PAE based on a basic model assumption, i.e., by dividing the BS array into multiple subarrays, the MS antennas can be considered to be in the FFR of each BS subarray. The details are introduced in the subsequent section.

{\color{black}
\begin{remark}
    In the proposed signal model, activated MS antennas can be assigned fully orthogonal time-frequency resources for transmission, rather than sharing the same resources for antennas with identical indices across different MS arrays, as currently proposed. This approach simplifies the position and attitude estimation by decoupling it into independent problems for each MS. Such a fully orthogonal transmission strategy, however, requires a larger amount of time-frequency resource.
\end{remark}

}


\section{Array Partitioning and Probabilistic System Model}

\subsection{Subarray-wise Far-Field Assumption}
\begin{figure}[t]
    \centering
    \begin{subfigure}[b]{.51\linewidth}
        \centering
        \includegraphics[width=.85\linewidth]{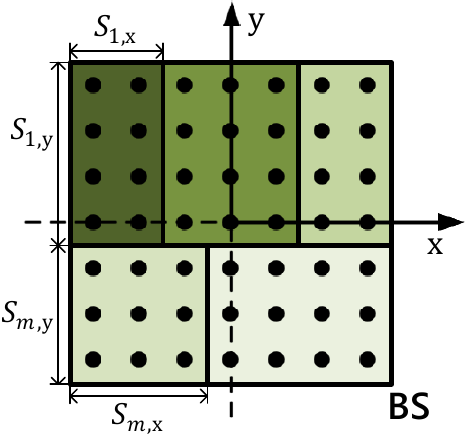}
        \caption{}
        \label{Apart}
    \end{subfigure}
    \begin{subfigure}[b]{.47\linewidth}
        \centering
        \includegraphics[width=.85\linewidth]{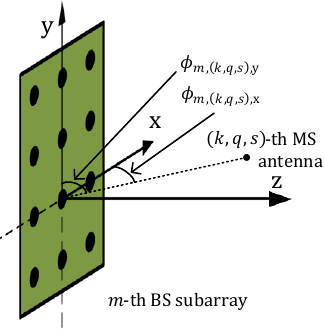}
        \caption{}
        \label{}
    \end{subfigure}
    \caption{(a) Illustration of the BS array partitioning. $M=5$. (b) Illustration of the AoAs at the $m$-th subarray.}
\end{figure}

We first give a simplified model of the system described in the previous section. The simplified system model is based on a basic assumption, that is, by dividing the BS array into multiple subarrays, all the MS antennas can be viewed as in the FFR of each subarray. Specifically, by following a partition strategy $\mathcal{D}$, the BS array can be partitioned into $M_{\mathcal{D}}$ non-overlapped rectangular subarrays, as shown in Fig. \ref{Apart}. The $m$-th subarray consists of $N_{m}=N_{m,\mathsf{x}} N_{m,\mathsf{y}}$ antennas, where $N_{m,\mathsf{x}}$ and $N_{m,\mathsf{y}}$ are the numbers of contained antennas along the $x$-axis and $y$-axis, respectively, $m\in \mathcal{I}_{M_{\mathsf{D}}}$. The size of the $m$-th subarray is given by $S_{m,\mathsf{x}}\times S_{m,\mathsf{y}}$. The largest dimension of the $m$-th subarray is $S_{m}\triangleq (S_{m,\mathsf{x}}^2 + S_{m,\mathsf{y}}^2)^{\frac{1}{2}}$, and the Rayleigh distance of the $m$-th subarray is given by $D_{\mathsf{R},m} \triangleq \frac{2S_{m}^2}{\lambda}$. Define a index transformation function $(m,\imath,\jmath)=\mathcal{G}_{\mathcal{D}}(u,v)$. The $(u,v)$-th BS antenna is indexed as the $(m,\imath,\jmath)$-th antenna in the $m$-th subarray, where $m\in \mathcal{I}_{M_{\mathcal{D}}}$, $\imath \in \mathcal{I}_{N_{m,\mathsf{x}}}$, $\jmath \in \mathcal{I}_{N_{m,\mathsf{y}}}$, $u\in \mathcal{I}_{N_{\mathsf{B};\mathsf{x}}}$, $v\in\mathcal{I}_{N_{\mathsf{B};\mathsf{y}}}$. For notational brevity, we henceforth omit the partition strategy subscript $\mathcal{D}$. Take the $(m,\tilde{N}_{m,\mathsf{x}}, \tilde{N}_{m,\mathsf{y}})$-th antenna as the reference antenna of the $m$-th BS subarray, where $\tilde{N}_{m,\mathsf{x}}\triangleq \lceil \frac{N_{m,\mathsf{x}}}{2} \rceil$, $\tilde{N}_{m,\mathsf{y}}\triangleq \lceil \frac{N_{m,\mathsf{y}}}{2} \rceil$, and $\lceil \cdot \rceil$ is the ceil operator, $m\in \mathcal{I}_{M}$. Denote by $\mathbf{p}_{\mathsf{B};m}$ the position of the reference antenna of the $m$-th subarray, $m\in \mathcal{I}_{M}$. The distance between the $(k,q,s)$-th MS antenna and the reference antenna of the $m$-th subarray is $r_{m,(k,q,s)}= \|\mathbf{p}_{\mathsf{M};(k,q,s)}-\mathbf{p}_{\mathsf{B};m}\|$.
The formal form of the SWFF assumption is given as follows.
\begin{assump}\label{SubArrAssump}
    With a sufficiently large $M$ and an appropriate partition strategy, all of the MS antennas can be viewed as in the FFR of each BS subarrays, i.e.,
    \begin{equation}
        \!\!\!\! D_{\mathsf{R},m} \! < \! r_{m,(k,q,s)},~ \forall m \! \in \! \mathcal{I}_{M}, \forall k \! \in \! \mathcal{I}_{K}, \forall q \! \in \! \mathcal{I}_{N_{\mathsf{M};\mathsf{x}}}, \forall s \!\in \! \mathcal{I}_{N_{\mathsf{M};\mathsf{y}}}.
    \end{equation}
\end{assump}
Assumption \ref{SubArrAssump} implies that by properly partitioning the BS array, the MS antennas can be viewed as in the NFR of the entire BS array, but in the FFR of each subarray. For a simple justification, we consider a BS array of the size $0.5\textrm{m}\times 0.5\textrm{m}$ and the carrier wavelength $\lambda=0.004$m. By letting $M=25$, the original BS array can be uniformly divided into subarrays on a $5\times 5$ grid, with each subarray of the same size $0.1\textrm{m}\times 0.1\textrm{m}$. The Rayleigh distance for each subarray is of $D_{\mathsf{R},m}=10$m. This means that Assumption \ref{SubArrAssump} holds when the distances between MS antennas and subarrays exceed 10m. In contrast, the Rayleigh distance for the entire BS array is $D_{\mathsf{R}}=250$m. There is a large range overlap between the FFR of each subarray and the NFR of the entire BS array. 

Based on \eqref{hcoeff}, the channel between the $(\imath,\jmath)$-th antenna of the $m$-th BS subarray and the $(k,q,s)$-th MS antenna can be expressed as 
\begin{equation}\label{hmijkqs0}
    h_{(m,\imath,\jmath),(k,q,s)} = \beta_k \rho_{(m,\imath,\jmath),(k,q,s)}e^{-\frac{j2\pi r_{(m,\imath,\jmath),(k,q,s)}}{\lambda}},
\end{equation}
where $\rho_{(m,\imath,\jmath),(k,q,s)}$ and $r_{(m,\imath,\jmath),(k,q,s)}$ are the path loss and the link distance, respectively, $m\in \mathcal{I}_M$, $\imath \in \mathcal{I}_{N_{m,\mathsf{x}}}$, $\jmath \in \mathcal{I}_{N_{m,\mathsf{y}}}$, $k\in \mathcal{I}_K$, $q\in \mathcal{I}_{N_{\mathsf{M};\mathsf{x}}}$, $s\in\mathcal{I}_{N_{\mathsf{M};\mathsf{y}}}$. Under Assumption \ref{SubArrAssump}, the channel representation \eqref{hmijkqs0} can be simplified by using the far-field channel model. Specifically, the path loss between an MS antenna and various antennas on a BS subarray can be approximated as the same, i.e.,
\begin{align}\label{rhomijkqs}
    \rho_{(m,\imath,\jmath),(k,q,s)} = \rho_{m,(k,q,s)},
\end{align}
where $\rho_{m,(k,q,s)} \triangleq \rho_{(m,\lceil \frac{N_{m,\mathsf{x}}}{2} \rceil, \lceil \frac{N_{m,\mathsf{y}}}{2} \rceil),(k,q,s)}$, 
$\forall \imath \in \mathcal{I}_{N_{m,\mathsf{x}}}$, $\forall \jmath \in \mathcal{I}_{N_{m,\mathsf{y}}}$, $m\in \mathcal{I}_M$, $k\in \mathcal{I}_K$, $q\in \mathcal{I}_{N_{\mathsf{M};\mathsf{x}}}$, $s\in\mathcal{I}_{N_{\mathsf{M};\mathsf{y}}}$. Such a common path loss approximation is widely used in the far-field channel models. Besides, the wavefront radiated by an MS antenna can be approximated as a plane when arriving at a BS subarray, resulting in linear phase difference of the channel coefficients at neighboring subarray antennas. 
Let $\phi_{m,(k,q,s),\mathsf{x}}\in (0,\pi)~ (\textrm{or}~ \phi_{m,(k,q,s),\mathsf{y}}\in (0,\pi))$ be the angle between $\mathbf{p}_{\mathsf{M};(k,q,s)}-\mathbf{p}_{\mathsf{B};m}$ and the $x$-axis (or $y$-axis). Define the cosine values of $\phi_{m,(k,q,s),\mathsf{x}}$ and $\phi_{m,(k,q,s),\mathsf{y}}$ as $\varphi_{m,(k,q,s),\ell} \triangleq \cos\phi_{m,(k,q,s),\ell}\in[-1,1]$, $\ell\in \{\mathsf{x},\mathsf{y}\}$.
$\varphi_{m,(k,q,s),\mathsf{x}}$ and $\varphi_{m,(k,q,s),\mathsf{y}}$ can also be expressed as the projections of the normalized $\mathbf{p}_{\mathsf{M};(k,q,s)}-\mathbf{p}_{\mathsf{B};m}$ on the $x$- and $y$- axes, respectively, i.e.,
\begin{align}
    \varphi_{m,(k,q,s),\ell} &= \frac{\left(\mathbf{p}_{\mathsf{M};(k,q,s)}-\mathbf{p}_{\mathsf{B};m}\right)^{\mathsf{T}}\boldsymbol{e}_{\ell}}{\|\mathbf{p}_{\mathsf{M};(k,q,s)}-\mathbf{p}_{\mathsf{B};m}\|}, ~ \ell\in\{\mathsf{x},\mathsf{y}\}. \label{varphix}
\end{align}
The positions of the $(m,\imath,\jmath)$-th BS antenna and the $(k,q,s)$-th MS antenna can be reformed as
\begin{align}
    &{\mathbf{p}}_{\mathsf{B};(m,\imath,\jmath)} \! = \! \mathbf{p}_{\mathsf{B};m} \! + \! \left[ \!\left( \! \imath \!- \! \tilde{N}_{m,\mathsf{x}} \! \right) \! \frac{\lambda}{2}, \! \left( \! \jmath \! - \! \tilde{N}_{m,\mathsf{y}} \! \right) \! \frac{\lambda}{2},0   \right]^{\mathsf{T}} \! \textrm{and}\\
    &{\mathbf{p}}_{\mathsf{M};(k,q,s)} = \mathbf{p}_{\mathsf{B};m} +  r_{m,(k,q,s)} \left[\varphi_{m,(k,q,s),\mathsf{x}}, \varphi_{m,(k,q,s),\mathsf{y}}, \right. \notag \\
    & \quad \quad \quad \quad \quad \quad \quad \left. \sqrt{1-\varphi_{m,(k,q,s),\mathsf{x}}^2 - \varphi_{m,(k,q,s),\mathsf{y}}^2}  \right]^{\mathsf{T}},
\end{align}
respectively, $m\in \mathcal{I}_M$, $\imath \in \mathcal{I}_{N_{m,\mathsf{x}}}$, $\jmath \in \mathcal{I}_{N_{m,\mathsf{y}}}$, $k\in \mathcal{I}_K$, $q\in \mathcal{I}_{N_{\mathsf{M};\mathsf{x}}}$, $s\in\mathcal{I}_{N_{\mathsf{M};\mathsf{y}}}$. 
Let $\imath_m \triangleq (\imath-\tilde{N}_{m,\mathsf{x}})$ and $\jmath_m \triangleq (\jmath-\tilde{N}_{m,\mathsf{y}})$, $\imath \in \mathcal{I}_{N_{m,\mathsf{x}}}$, $\jmath \in \mathcal{I}_{N_{m,\mathsf{y}}}$.
Under the far-field channel model, the link distance $r_{(m,\imath,\jmath),(k,q,s)}$ in \eqref{hmijkqs0} can be expressed as 
\begin{subequations}\label{rmijkqs}
    \begin{align} 
        r_{(m,\imath,\jmath),(k,q,s)} & \!=\!  \|{\mathbf{p}}_{\mathsf{M};(k,q,s)} \!- \!{\mathbf{p}}_{\mathsf{B};(m,\imath,\jmath)}\|, \label{rmijkqs1} \\
        & \!=\! r_{m,(k,q,s)} \! - \! \frac{\lambda \varphi_{m,(k,q,s),\mathsf{x}} \imath_m}{2} \! - \! \frac{\lambda \varphi_{m,(k,q,s),\mathsf{y}} \jmath_m}{2} \label{rmijkqs3},
    \end{align}
\end{subequations}
where \eqref{rmijkqs3} is the first-order Taylor expansion of \eqref{rmijkqs1} at $\frac{\lambda}{2r_{m,(k,q,s)}}=0$, $m\in \mathcal{I}_M$, $\imath \in \mathcal{I}_{N_{m,\mathsf{x}}}$, $\jmath \in \mathcal{I}_{N_{m,\mathsf{y}}}$, $k\in \mathcal{I}_K$, $q\in \mathcal{I}_{N_{\mathsf{M};\mathsf{x}}}$, $s\in\mathcal{I}_{N_{\mathsf{M};\mathsf{y}}}$. Based on \eqref{rhomijkqs} and \eqref{rmijkqs}, under Assumption 1, the channel model \eqref{hmijkqs0} is reduced to 
\begin{align}\label{hReduced}
    h_{(m,\imath,\jmath),(k,q,s)} &= \beta_k \rho_{m,(k,q,s)} e^{-j\psi_{(m,\imath,\jmath),(k,q,s)}},~ \textrm{where} \\
    \psi_{(m,\imath,\jmath),(k,q,s)} & \!= \!  \frac{2\pi r_{m,(k,q,s)}}{\lambda}
    \!+\! \pi  \left(\imath-\tilde{N}_{m,\mathsf{x}}\right) \varphi_{m,(k,q,s),\mathsf{x}} \notag \\
    &\quad + \left(\jmath-\tilde{N}_{m,\mathsf{y}} \right) \varphi_{m,(k,q,s),\mathsf{y}}. \label{psimijkqs}
\end{align}
Based on the SWFF channel model \eqref{hReduced}, and following the signal transmission strategy described in Sec. \ref{secSigModel}, the baseband received signal model at the $m$-th BS subarray in the $t$-th time slot is given by 
\begin{equation}\label{Ymt}
    \boldsymbol{Y}_{m,t}  =  \sum_{k\in\mathcal{I}_K}  \varrho_{m,(k,t)} \boldsymbol{\varUpsilon}_{m,(k,t)}\left(\boldsymbol{\varphi}_{m,(k,t)}\right) + \boldsymbol{w}_{m,t},
\end{equation} 
where
\begin{multline}\label{alphamkt}
    \varrho_{m,(k,t)}\triangleq x_{k,t} \beta_k \rho_{m,(k,t)} e^{-\frac{j2\pi r_{m,(k,t)}}{\lambda}} \\
    \times e^{j\pi\left(\tilde{N}_{m,\mathsf{x}}\varphi_{m,(k,t),\mathsf{x}} + \tilde{N}_{m,\mathsf{y}}\varphi_{m,(k,t),\mathsf{y}}\right)}
\end{multline}
is a complex coefficient irrelevant to the subarray antenna indices $\imath$ and $\jmath$, and the $(\imath,\jmath)$-th element of $\boldsymbol{\varUpsilon}_{m,(k,t)}\left(\boldsymbol{\varphi}_{m,(k,t)}\right) \in \mathbb{C}^{N_{m,\mathsf{x}}\times N_{m,\mathsf{y}}}$ is given by 
\begin{equation}
    \varUpsilon_{(m,\imath,\jmath),(k,t)} \triangleq e^{-j\pi\left( \imath \varphi_{m,(k,t),\mathsf{x}} + \jmath \varphi_{m,(k,t),\mathsf{y}} \right)},
\end{equation}
$\imath \in \mathcal{I}_{N_{m,\mathsf{x}}}$, $\jmath \in \mathcal{I}_{N_{m,\mathsf{y}}}$.
The received signals at various subarrays can be assembled to the received signal at the entire BS array. Define a matrix $\boldsymbol{Y}_{t}\in \mathbb{C}^{N_{\mathsf{B};\mathsf{x}}\times N_{\mathsf{B};\mathsf{y}}}$, where the $(u,v)$-th element of $\boldsymbol{Y}_{t}$ is given by the $(m,\imath,\jmath)$-th element of $\boldsymbol{Y}_{m,t}$, with $(m,\imath,\jmath)=\mathcal{G}(u,v)$, $m\in \mathcal{I}_M$, $\imath \in \mathcal{I}_{N_{m,\mathsf{x}}}$, $\jmath \in \mathcal{I}_{N_{m,\mathsf{y}}}$, $u\in \mathcal{I}_{N_{\mathsf{B};\mathsf{x}}}$, $v\in\mathcal{I}_{N_{\mathsf{B};\mathsf{y}}}$. Then, the received signals in the entire $T$ time slots can be written as 
\begin{equation}\label{SubFarSigMod}
    \boldsymbol{Y} = \left[ \textrm{vec}(\boldsymbol{Y}_1),\ldots,\textrm{vec}(\boldsymbol{Y}_{T})  \right]\in \mathbb{C}^{N_{\mathsf{B}}\times T}.
\end{equation}
Eqn. \eqref{SubFarSigMod} is referred to as the \textit{SWFF} received signal model, which is a simplified version of the original near-field signal model \eqref{Y} based on Assumption 1. 

{\color{black}
\begin{remark}
    For the proposed localization framework, the signals transmitted by the MSs can be communication signals carrying data, and are not necessarily to be pilot signals known in advance at the BS. As shown in \eqref{alphamkt}, the signal transmitted from the MS, $x_{k,t}$, is absorbed into a complex coefficient $\varrho_{m,(k,t)}$. Regardless of whether $x_{k,t}$ is known in advance, $\varrho_{m,(k,t)}$ is considered unknown and serves as a nuisance parameter in the proposed method. This complex coefficient is estimated alongside with the AoAs in the AoA estimation module (as detailed later in Sec. IV. A).
\end{remark}
}

\subsection{Probabilistic Problem Formulation}

We now establish the probability model of the considered PAE problem under the SWFF signal model. The complex channel gain $\varrho_{m,(k,t)}$ in \eqref{alphamkt} is assigned a CSCG prior, i.e., 
\begin{equation}\label{pvarrho}
    p(\varrho_{m,(k,t)}) \sim  \mathcal{CN}(0,\sigma_{\varrho}^2), m\in \mathcal{I}_{M}, k\in \mathcal{I}_{K}, t\in \mathcal{I}_{T}.
\end{equation}
The central position of the $k$-th MS array is assigned a zero-mean Gaussian prior, i.e.,
\begin{equation}\label{ppMk}
    p(\mathbf{p}_{\mathsf{M};k}) = \mathcal{N}(\boldsymbol{0},\sigma_{\mathbf{p};k}^2 \boldsymbol{I}), ~k\in \mathcal{I}_{K}.
\end{equation}
The rotation angles of the $k$-th MS array follow Von-Mises (VM) priors as 
\begin{align}
    p(\theta_{\mathsf{x};k}) &= \mathcal{M}(\theta_{\mathsf{x};k};\chi_{\mathsf{x};k},\kappa_{\mathsf{x};k}), \\
    p(\theta_{\mathsf{y};k}) &= \mathcal{M}(2\theta_{\mathsf{y};k};\chi_{\mathsf{y};k},\kappa_{\mathsf{y};k}),~ \textrm{and} \label{VMyk}\\
    p(\theta_{\mathsf{z};k}) &= \mathcal{M}(\theta_{\mathsf{z};k};\chi_{\mathsf{z};k},\kappa_{\mathsf{z};k}),~ \textrm{where} \\
    \mathcal{M}(\vartheta;\chi,\kappa) &= \frac{\exp\left( \kappa \cos(\vartheta - \chi) \right)}{2\pi I_{0}(\kappa)}, k\in \mathcal{I}_{K} \label{VMpdf}
\end{align}
In \eqref{VMpdf}, $I_0$ represents the zero-order modified Bessel function of the first kind; $\chi$ is the mean direction, and $\kappa>0$ is the concentration parameter. {\color{black} The VM distribution is widely used in statistical inference to model angular random variables defined on the unit circle. It shares several properties with the Gaussian distribution, such as the maximum entropy property and closure under multiplication \cite{jammalamadaka2001topics}.} In this paper, the angle support of $\vartheta$ is specified to $[-\pi,\pi]$. A large $\kappa$ means that the distribution is concentrated around the mean direction $\chi$, while a small $\kappa$ means the distribution is close to uniform. The coefficient $2$ in \eqref{VMyk} ensures the standard form of the VM distribution for $\theta_{\mathsf{y};k}\in[-\frac{\pi}{2},\frac{\pi}{2})$. The prior distribution of $\boldsymbol{\theta}_k$ is then given by 
\begin{equation}\label{pthetak}
    p(\boldsymbol{\theta}_k) = p(\theta_{\mathsf{x};k}) p(\theta_{\mathsf{y};k}) p(\theta_{\mathsf{z};k}).
\end{equation}
When there is no prior knowledge of the MSs' positions and attitudes, $\sigma_{\mathbf{p};k}$, $\kappa_{\mathsf{x};k}$, $\kappa_{\mathsf{y};k}$, and $\kappa_{\mathsf{z};k}$ can be set to large values to make the prior distributions non-informative, $\forall k \in \mathcal{I}_{K}$.

Based on \eqref{pMkqs}, the probability distribution of $\mathbf{p}_{\mathsf{M};(k,t)}$ conditioned on $\mathbf{p}_{\mathsf{M};k}$ and $\boldsymbol{\theta}_k$ is given by 
\begin{equation}\label{ppMpMkthetak}
    p\left(\mathbf{p}_{\mathsf{M};(k,t)}|\mathbf{p}_{\mathsf{M};k}, \boldsymbol{\theta}_k  \right)  =  \delta  \left( \mathbf{p}_{\mathsf{M};(k,t)}  -  \mathbf{p}_{\mathsf{M};k}  -  \mathbf{R}(\boldsymbol{\theta}_k) \mathbf{q}_{\mathsf{M};(k,t)}  \right) . 
\end{equation}
Based on \eqref{varphix}, the distribution of $\boldsymbol{\varphi}_{m,(k,t)}$ conditioned on $\mathbf{p}_{\mathsf{M};(k,t)}$ can be expressed as 
\begin{align}
    &p(\boldsymbol{\varphi}_{m,(k,t)}|\mathbf{p}_{\mathsf{M};(k,t)}) \notag \\
    &= \prod_{\ell\in\{\mathsf{x},\mathsf{y}\}} \delta \left(\varphi_{m,(k,t),\ell} - \frac{\left(\mathbf{p}_{\mathsf{M};(k,t)}-\mathbf{p}_{\mathsf{B};m}\right)^{\mathsf{T}}\boldsymbol{e}_{\ell}}{\|\mathbf{p}_{\mathsf{M};(k,t)}-\mathbf{p}_{\mathsf{B};m}\|}\right). \label{pVarphipM} 
\end{align}
Based on \eqref{Ymt}, the likelihood function of the received signal at the $m$-th subarray in the $t$-th time slot is given by 
\begin{align}
    & p\left(\boldsymbol{Y}_{m,t} \Big|\left\{\varrho_{m,(k,t)},\boldsymbol{\varphi}_{m,(k,t)}|k\in\mathcal{I}_K\right\}\right) \notag \\
    &= \mathcal{CN}\left(\!\textrm{vec}\left(\boldsymbol{Y}_{m,t} \!- \! \sum_{k\in\mathcal{I}_K} \varrho_{m,(k,t)} \boldsymbol{\varUpsilon}_{m,(k,t)}\right), \sigma_{w}^2 \boldsymbol{I} \! \right). \label{pYmtlikelihood}
\end{align}
Define 
\begin{align}
    \mathcal{S}_{\varrho} &\triangleq \left\{\varrho_{m,(k,t)}|m\in\mathcal{I}_{M},k\in\mathcal{I}_{K},t\in\mathcal{I}_{T}\right\}, \\
    \mathcal{S}_{\boldsymbol{\varphi}} &\triangleq \left\{\boldsymbol{\varphi}_{m,(k,t)}|m\in\mathcal{I}_{M},k\in\mathcal{I}_{K},t\in\mathcal{I}_{T}\right\}, \\
    \mathcal{S}_{\mathbf{p}_{\mathsf{A}}} &\triangleq \left\{\mathbf{p}_{\mathsf{M};(k,t)}|m\in\mathcal{I}_{M},k\in\mathcal{I}_{K},t\in\mathcal{I}_{T}\right\}, \\
    \mathcal{S}_{\mathbf{p}_{\mathsf{M}}} &\triangleq \left\{\mathbf{p}_{\mathsf{M};k}|k\in\mathcal{I}_{K}\right\}, ~\mathcal{S}_{\boldsymbol{\theta}} \triangleq \left\{\boldsymbol{\theta}_{k}|k\in\mathcal{I}_{K}\right\},\\
    \mathcal{S}_{\mathbf{p}_{\mathsf{M}}\backslash k} &\triangleq \left\{\mathbf{p}_{\mathsf{M};k'}|k'= 1,\ldots,k-1,k+1,\ldots,K\right\}, \\
    \textrm{and}~ \mathcal{S}_{\boldsymbol{\theta} \backslash k} &\triangleq \left\{\boldsymbol{\theta}_{k'}|k'=1,\ldots,k-1,k+1,\ldots,K\right\},
\end{align}
{\color{black}as the sets of complex attitudes, AoAs, positions of the activated MS antennas, positions of the MSs, rotation angles of the MSs, positions of the MSs except the $k$-th one, rotation angles of the MSs except the $k$-th one, respectively.}
{\color{black}Based on the dependencies among these variables shown in \eqref{ppMpMkthetak}-\eqref{pYmtlikelihood}, the joint distribution of $\boldsymbol{Y}$, $\mathcal{S}_{\varrho}$, $\mathcal{S}_{\boldsymbol{\varphi}}$, $\mathcal{S}_{\mathbf{p}_{\mathsf{A}}}$, $\mathcal{S}_{\mathbf{p}_{\mathsf{M}}}$, and $\mathcal{S}_{\theta}$ is }
\begin{align}
    &p(\boldsymbol{Y},\mathcal{S}_{\varrho},\mathcal{S}_{\boldsymbol{\varphi}},\mathcal{S}_{\mathbf{p}_{\mathsf{A}}},\mathcal{S}_{\mathbf{p}_{\mathsf{M}}},\mathcal{S}_{\theta}) \notag \\
    &= \prod_{t\in \mathcal{I}_T} \prod_{m\in \mathcal{I}_M} p\left(\boldsymbol{Y}_{m,t} \Big|\left\{\varrho_{m,(k,t)},\boldsymbol{\varphi}_{m,(k,t)}|k\in\mathcal{I}_K\right\}\right) \notag \\
    &\quad \times \prod_{k\in \mathcal{I}_K} p(\varrho_{m,(k,t)}) p(\boldsymbol{\varphi}_{m,(k,t)}|\mathbf{p}_{\mathsf{M};(k,t)})  \notag \\
    &\quad \times p\left(\mathbf{p}_{\mathsf{M};(k,t)}|\mathbf{p}_{\mathsf{M};k},\boldsymbol{\theta}_{k}  \right) p(\mathbf{p}_{\mathsf{M};k}) p(\boldsymbol{\theta}_k). \label{JointPDF}
\end{align}
{\color{black}From Bayes' theorem, the posterior distributions of $\boldsymbol{\theta}_k$ and $\mathbf{p}_{\mathsf{M};k}$ can be obtained by marginalizing the joint posterior distribution $p(\mathcal{S}_{\varrho},\mathcal{S}_{\boldsymbol{\varphi}},\mathcal{S}_{\mathbf{p}_{\mathsf{A}}},\mathcal{S}_{\mathbf{p}_{\mathsf{M}}},\mathcal{S}_{\theta}|\boldsymbol{Y})$ over all variables except the ones of interest, as given by}
\begin{align}
    &p(\boldsymbol{\theta}_k |\boldsymbol{Y})\\
    &= \int \frac{p(\boldsymbol{Y},\mathcal{S}_{\varrho},\mathcal{S}_{\boldsymbol{\varphi}},\mathcal{S}_{\mathbf{p}_{\mathsf{A}}},\mathcal{S}_{\mathbf{p}_{\mathsf{M}}},\mathcal{S}_{\boldsymbol{\theta}})}{p(\boldsymbol{Y})} \textrm{d}\mathcal{S}_{\varrho} \textrm{d} \mathcal{S}_{\boldsymbol{\varphi}} \textrm{d} \mathcal{S}_{\mathbf{p}_{\mathsf{A}}} \textrm{d} \mathcal{S}_{\mathbf{p}_{\mathsf{M}}} \textrm{d} \mathcal{S}_{\boldsymbol{\theta}\backslash k},  \notag  \\ 
    &p(\mathbf{p}_{\mathsf{M};k}| \boldsymbol{Y})  \\
    &= \int \frac{p(\boldsymbol{Y},\mathcal{S}_{\varrho},\mathcal{S}_{\boldsymbol{\varphi}},\mathcal{S}_{\mathbf{p}_{\mathsf{A}}},\mathcal{S}_{\mathbf{p}_{\mathsf{M}}},\mathcal{S}_{\boldsymbol{\theta}})}{p(\boldsymbol{Y})} \textrm{d}\mathcal{S}_{\varrho} \textrm{d} \mathcal{S}_{\boldsymbol{\varphi}} \textrm{d} \mathcal{S}_{\mathbf{p}_{\mathsf{A}}} \textrm{d} \mathcal{S}_{\boldsymbol{\theta}} \textrm{d} \mathcal{S}_{\mathbf{p}_{\mathsf{M}}\backslash k}. \notag
\end{align}
An estimate of $\boldsymbol{\theta}_k$ and $\mathbf{p}_{\mathsf{M};k}$ can be obtained by following the minimum mean-squared error (MMSE) or maximum \textit{a posteriori} (MAP) criterions. Exact MMSE or MAP estimation is often impractical due to the high computational complexity for the posterior distributions. Therefore, we propose a low-complexity approach based on message passing, i.e., the APPLE algorithm, as explained in the following section.





\section{Proposed APPLE Algorithm}

\begin{figure*}[t]
    \centering
    \begin{subfigure}[b]{.6\linewidth}
        \centering
        \includegraphics[height=5.5cm]{FactorGraph.pdf}
    \end{subfigure}
    \begin{subfigure}[b]{.39\linewidth}
        \centering
        \includegraphics[height=5.3cm]{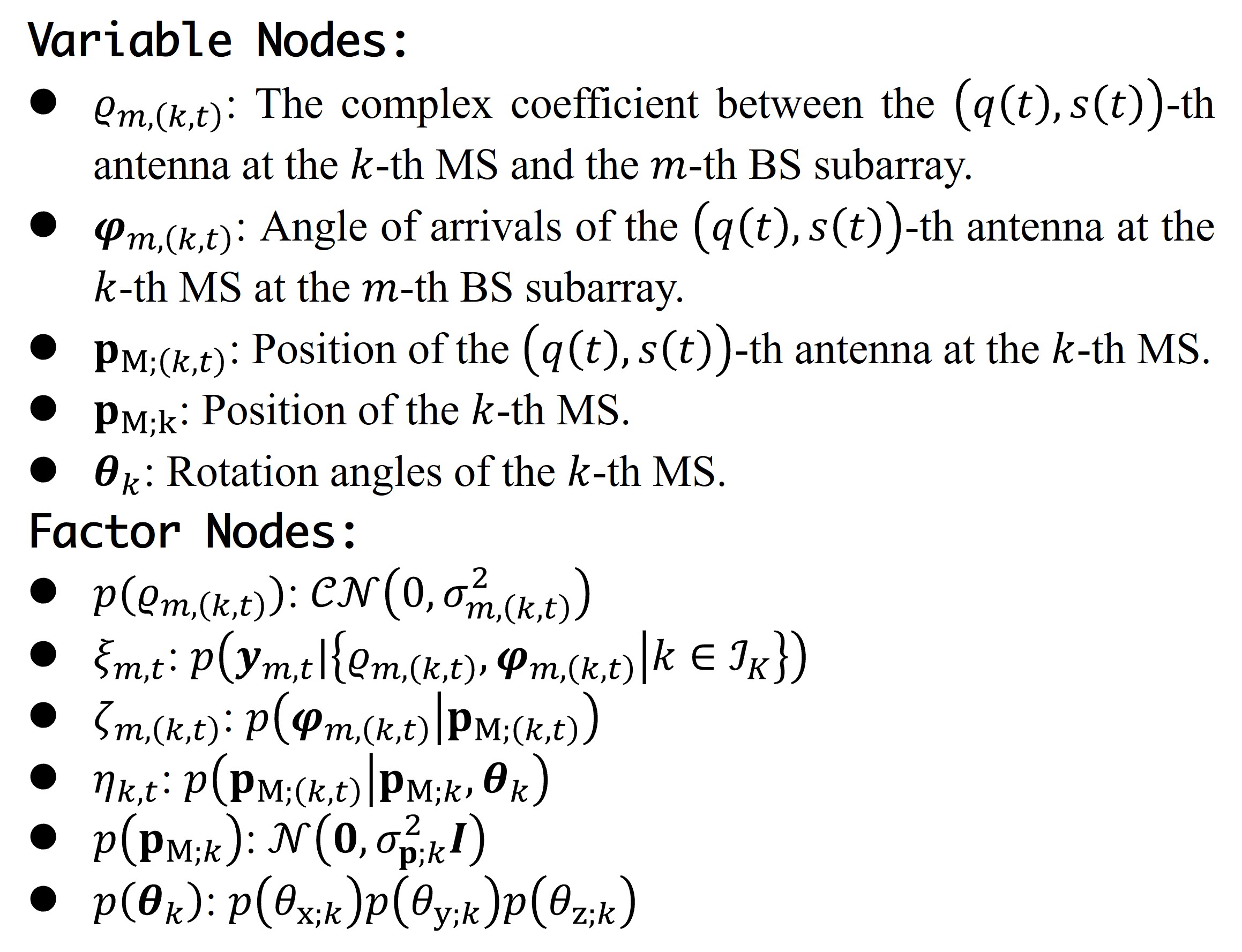}
    \end{subfigure}
    \caption{Factor graph of the probability model.}
    \label{FacGraph}
\end{figure*}
This section presents the proposed message passing based APPLE algorithm. The factor graph representation of the probability model \eqref{JointPDF} is shown in Fig. \ref{FacGraph}, where circles represent variable nodes and squares represent factor nodes. The factor graph is divided into two modules, namely, the AoA estimation module that obtains the AoA estimates of the activated MS antennas at various BS subarrays, and the information fusion module that combines the AoA estimates and the prior topological knowledge of the activated MS antennas. For notational brevity, we denote by $\xi_{m,t}$ the factor node $p\left(\boldsymbol{Y}_{m,t} \Big|\left\{\varrho_{m,(k,t)},\boldsymbol{\varphi}_{m,(k,t)}|k\in\mathcal{I}_K\right\}\right)$, by $\zeta_{m,(k,t)}$ the factor node $p(\boldsymbol{\varphi}_{m,(k,t)}|\mathbf{p}_{\mathsf{M};(k,t)})$, and by $\eta_{k,t}$ the factor node $p\left(\mathbf{p}_{\mathsf{M};(k,t)}|\mathbf{p}_{\mathsf{M};k},\boldsymbol{\theta}_k  \right)$. Denote by $\Delta_{g\rightarrow h}$ the message from node $g$ to node $h$, and by $\boldsymbol{n}_{g\rightarrow h}$ and $\boldsymbol{C}_{g\rightarrow h}$ the mean and covariance matrix of message $\Delta_{g\rightarrow h}$, respectively.

\subsection{AoA Estimation Module}
We first consider message passing in the AoA estimation module. We start with the messages input into the AoA estimation module. As shown in the subsequent subsection, the message from the variable node $\mathbf{p}_{\mathsf{M};(k,t)}$ to the factor node $\zeta_{m,(k,t)}$ is a Gaussian distribution, given by 
\begin{multline}\label{Deltapm2zeta}
    \Delta_{\mathbf{p}_{\mathsf{M};(k,t)} \rightarrow \zeta_{m,(k,t)}} \\
     = \mathcal{N}(\mathbf{p}_{\mathsf{M};(k,t)};\boldsymbol{n}_{\mathbf{p}_{\mathsf{M};(k,t)} \rightarrow \zeta_{m,(k,t)}} ,\boldsymbol{C}_{\mathbf{p}_{\mathsf{M};(k,t)} \rightarrow \zeta_{m,(k,t)}}),
\end{multline}
where $\boldsymbol{n}_{\mathbf{p}_{\mathsf{M};(k,t)} \rightarrow \zeta_{m,(k,t)}}$ and $\boldsymbol{C}_{\mathbf{p}_{\mathsf{M};(k,t)} \rightarrow \zeta_{m,(k,t)}}$ are the mean and the covariance matrix, respectively. 
\subsubsection{Messages from $\zeta_{m,(k,t)}$ to $\boldsymbol{\varphi}_{m,(k,t)}$}
Following the sum-product rule, the message from the factor node $\zeta_{m,(k,t)}$ to the variable node $\boldsymbol{\varphi}_{m,(k,t)}$ is given by 
\begin{equation}\label{Deltazeta2varphi}
    \Delta_{\zeta_{m,(k,t)} \rightarrow \boldsymbol{\varphi}_{m,(k,t)}} \! \propto \! \int_{\mathbf{p}_{\mathsf{M};(k,t)}} \!\!\!\!\!\!\!\!\!\!\! p(\boldsymbol{\varphi}_{m,(k,t)}|\mathbf{p}_{\mathsf{M};(k,t)}) \Delta_{\mathbf{p}_{\mathsf{M};(k,t)} \rightarrow \zeta_{m,(k,t)}}.
\end{equation}
The integral in \eqref{Deltazeta2varphi} generally has no closed-form expression. To simplify the subsequent message calculation, by following the method provided in \cite[Eqn. (39)]{tengBayesianUserLocalization2022}, we approximate this message by the product of two VM distributions as 
\begin{equation}\label{Gauss2VM}
    \Delta_{\zeta_{m,(k,t)} \rightarrow \boldsymbol{\varphi}_{m,(k,t)}}   = \!\!\! \prod_{\ell \in\{\mathsf{x},\mathsf{y}\}} \!\!\! \mathcal{M}(\pi\varphi_{m,(k,t),\mathsf{\ell}};\chi_{m,(k,t),\mathsf{x}}^{\mathsf{pri}},\kappa_{m,(k,t),\mathsf{\ell}}^{\mathsf{pri}}) 
\end{equation}
with
\begin{align}
    &\chi_{m,(k,t),\ell}^{\mathsf{pri}} = \pi \bar{\varphi}_{m,(k,t),\ell}, \label{chipri} \\
    &\kappa_{m,(k,t),\ell}^{\mathsf{pri}} \label{kappapri} \\
    &= \frac{d_{m,(k,t)}^2}{\pi^2\left(1-\bar{\varphi}_{m,(k,t),\ell}^2\right)\boldsymbol{v}_{m,(k,t),\ell}^{\mathsf{T}}\boldsymbol{C}_{\mathbf{p}_{\mathsf{M};(k,t)} \rightarrow \zeta_{m,(k,t)}} \boldsymbol{v}_{m,(k,t),\ell}^{\mathsf{T}}}.  \notag
\end{align}
In \eqref{kappapri}, $\boldsymbol{v}_{m,(k,t),\ell}$ is the unit direction vector perpendicular to $\left(\mathbf{p}_{\mathsf{B};m}-\boldsymbol{n}_{\mathbf{p}_{\mathsf{M};(k,t)} \rightarrow \zeta_{m,(k,t)}}\right)$ and is on the plane spanned by $\boldsymbol{e}_{\ell}$ and $\left(\mathbf{p}_{\mathsf{B};m}-\boldsymbol{n}_{\mathbf{p}_{\mathsf{M};(k,t)} \rightarrow \zeta_{m,(k,t)}}\right)$;
$\bar{\varphi}_{m,(k,t),\ell}$ is the mean AoA given by 
\begin{equation}
    \bar{\varphi}_{m,(k,t),{\ell}} = \frac{\left(\boldsymbol{n}_{\mathbf{p}_{\mathsf{M};(k,t)} \rightarrow \zeta_{m,(k,t)}}-\mathbf{p}_{\mathsf{B};m}\right)^{\mathsf{T}} \boldsymbol{e}_{{\ell}}}{\Big\|\boldsymbol{n}_{\mathbf{p}_{\mathsf{M};(k,t)} \rightarrow \zeta_{m,(k,t)}}-\mathbf{p}_{\mathsf{B};m}\Big\|}.
\end{equation}
$d_{m,(k,t)}$ is the distance between $\boldsymbol{n}_{\mathbf{p}_{\mathsf{M};(k,t)} \rightarrow \zeta_{m,(k,t)}}$ and $\mathbf{p}_{\mathsf{B};m}$ given by $d_{m,(k,t)} = \Big\|\boldsymbol{n}_{\mathbf{p}_{\mathsf{M};(k,t)} \rightarrow \zeta_{m,(k,t)}}-\mathbf{p}_{\mathsf{B};m}\Big\|$.
{\color{black}
The approximation in \eqref{Gauss2VM} is inspired by the mean field approximation in variational inference, where a joint posterior distribution is approximated by the product of multiple marginal posterior distributions \cite{badiuVariationalBayesianInference2017}. Approximating $\Delta_{\zeta_{m,(k,t)} \rightarrow \varphi_{m,(k,t),\mathsf{x}}}$ and $\Delta_{\zeta_{m,(k,t)} \rightarrow \varphi_{m,(k,t),\mathsf{y}}}$ as VM distributions is based on their extensive application in modeling angular variables.}
The detailed derivation of \eqref{Gauss2VM} can be found in \cite[Appendix B]{tengBayesianUserLocalization2022}.

\subsubsection{Messages from $\xi_{m,t}$ to $\boldsymbol{\varphi}_{m,(k,t)}$}
The message from the variable node $\boldsymbol{\varphi}_{m,(k,t)}$ to the factor node $\xi_{m,t}$ is equal to that from $\zeta_{m,(k,t)}$ to $\boldsymbol{\varphi}_{m,(k,t)}$, i.e., $\Delta_{\boldsymbol{\varphi}_{m,(k,t)}\rightarrow \xi_{m,t}} = \Delta_{\zeta_{m,(k,t)} \rightarrow \boldsymbol{\varphi}_{m,(k,t)}}$.
Then, the message from the factor node $\xi_{m,t}$ back to the variable node $\boldsymbol{\varphi}_{m,(k,t)}$ is given by 
\begin{align}
    &\Delta_{\xi_{m,t}\rightarrow \boldsymbol{\varphi}_{m,(k,t)}} \notag \\
    &\propto \int_{\mathcal{S}_{\varrho_{m,t}}}\int_{\mathcal{S}_{\boldsymbol{\varphi}_{m,(k,t)}}} p\left(\boldsymbol{Y}_{m,t} \Big|\left\{\varrho_{m,(k,t)},\boldsymbol{\varphi}_{m,(k,t)}|k\in\mathcal{I}_K\right\}\right) \notag \\
    &\quad\times p(\varrho_{m,(k,t)}) \! \prod_{k'\in\{\mathcal{I}_K\backslash k\}} p(\varrho_{m,(k',t)}) \Delta_{\boldsymbol{\varphi}_{m,(k',t)}\rightarrow \xi_{m,t}} , \label{xi2varphi}
\end{align}
where $\mathcal{S}_{\varrho_{m,t}} \triangleq \left\{ \varrho_{m,(k,t)}|k\in \mathcal{I}_K \right\}$ and $\mathcal{S}_{\boldsymbol{\varphi}_{m,(k,t)}} \triangleq \left\{ \boldsymbol{\varphi}_{m,(k',t)}|k'\in\{\mathcal{I}_K \backslash k\} \right\}$. Directly solving the integral on the RHS of \eqref{xi2varphi} is computationally prohibitive, and we introduce an approximate method for calculating $\Delta_{\xi_{m,t}\rightarrow \boldsymbol{\varphi}_{m,(k,t)}}$. {\color{black}Similar approximate message computation methods have been used in \cite{tengBayesianUserLocalization2022,zhengScalableNearFieldLocalization2023}.
Note that $\Delta_{\xi_{m,t}\rightarrow \boldsymbol{\varphi}_{m,(k,t)}}$ represents the belief of $\boldsymbol{\varphi}_{m,(k,t)}$ at the factor node $\xi_{m,t}$ based on messages from the other connected variable nodes except $\boldsymbol{\varphi}_{m,(k,t)}$. $\Delta_{\boldsymbol{\varphi}_{m,(k,t)}\rightarrow \xi_{m,t}}$ can be viewed as a prior distribution of $\boldsymbol{\varphi}_{m,(k,t)}$. 
From the perspective of sum-product rule \cite{kschischangFactorGraphsSumproduct2001}, the product of $\Delta_{\boldsymbol{\varphi}_{m,(k,t)}\rightarrow \xi_{m,t}} = \Delta_{\zeta_{m,(k,t)} \rightarrow \boldsymbol{\varphi}_{m,(k,t)}}$ and $\Delta_{\xi_{m,t}\rightarrow \boldsymbol{\varphi}_{m,(k,t)}}$ can be viewed as an approximate posterior distribution of $\boldsymbol{\varphi}_{m,(k,t)}$, i.e., 
\begin{equation}\label{VarphiPosterior}
    \Delta_{\xi_{m,t}\rightarrow \boldsymbol{\varphi}_{m,(k,t)}} \Delta_{\boldsymbol{\varphi}_{m,(k,t)}\rightarrow \xi_{m,t}} \propto p(\boldsymbol{\varphi}_{m,(k,t)}|\boldsymbol{Y}_{m,t}).
\end{equation}}
Based on \eqref{xi2varphi} and \eqref{VarphiPosterior}, $\Delta_{\xi_{m,t}\rightarrow \boldsymbol{\varphi}_{m,(k,t)}}$ can be rewritten as 
\begin{equation}\label{xi2varphi3}
    \Delta_{\xi_{m,t}\rightarrow \boldsymbol{\varphi}_{m,(k,t)}} \propto \frac{p(\boldsymbol{\varphi}_{m,(k,t)}|\boldsymbol{Y}_{m,t})}{\Delta_{\boldsymbol{\varphi}_{m,(k,t)}\rightarrow \xi_{m,t}}}.
\end{equation}
Calculating the posterior distribution $p(\boldsymbol{\varphi}_{m,(k,t)}|\boldsymbol{Y}_{m,t})$ based on the prior $\Delta_{\boldsymbol{\varphi}_{m,(k,t)}\rightarrow \xi_{m,t}}$ is actually a Bayesian line spectra estimation problem, and can be efficiently solved by using the VALSE algorithm provided in \cite{badiuVariationalBayesianInference2017}. The VALSE algorithm can efficiently compute the approximate posterior estimates of AoAs $\boldsymbol{\varphi}_{m,(k,t)}$ and the complex coefficient $\varrho_{m,(k,t)}$ by following the variational inference principle. {\color{black}Note that the VALSE algorithm also outputs the posterior distribution of the complex coefficient $\varrho_{m,(k,t)}$ at each BS subarray. These distributions of $\varrho_{m,(k,t)}$ are not utilized in later stages, as the complex coefficients are treated as nuisance parameters in our approach.} The returned posterior distribution $p(\boldsymbol{\varphi}_{m,(k,t)}|\boldsymbol{Y}_{m,t})$ by VALSE is also in the form of the product of two VM distributions with respect to $\varphi_{m,(k,t),\mathsf{x}}$ and $\varphi_{m,(k,t),\mathsf{y}}$, i.e.,
\begin{equation}
    p(\boldsymbol{\varphi}_{m,(k,t)}|\boldsymbol{Y}_{m,t})  = \!\! \prod_{\ell\in \{\mathsf{x},\mathsf{y}\}} \!\!\!\! \mathcal{M}(\pi\varphi_{m,(k,t),\ell};\chi_{m,(k,t),\ell}^{\mathsf{post}},\kappa_{m,(k,t),\ell}^{\mathsf{post}}) 
\end{equation}
Then, based on \eqref{xi2varphi3} and the closure of the VM distributions under multiplication\footnotemark, we have 
\footnotetext{The product of two VM probability density functions (pdfs) is proportional to another VM pdf, i.e.,
    $\mathcal{M}(\vartheta;\chi_1,\kappa_1) \mathcal{M}(\vartheta;\chi_2,\kappa_2) \propto \mathcal{M}(\vartheta;\chi_3,\kappa_3)$,
where $\kappa_3   =   |\kappa_1 e^{j\chi_1}   +   \kappa_2 e^{j\chi_2}|$ and $\chi_3   =   \angle (\kappa_1 e^{j\chi_1}   +   \kappa_2 e^{j\chi_2})$ \cite{jammalamadaka2001topics}. }
\begin{align}
    \Delta_{\xi_{m,t}\rightarrow \boldsymbol{\varphi}_{m,(k,t)}} &\propto \prod_{\ell\in \{\mathsf{x},\mathsf{y}\}} \frac{ \mathcal{M}(\pi\varphi_{m,(k,t),\ell};\chi_{m,(k,t),\ell}^{\mathsf{post}},\kappa_{m,(k,t),\ell}^{\mathsf{post}}) }{ \mathcal{M}(\pi\varphi_{m,(k,t),\ell};\chi_{m,(k,t),\ell}^{\mathsf{pri}},\kappa_{m,(k,t),\ell}^{\mathsf{pri}}) } \\
    &\propto \prod_{\ell\in \{\mathsf{x},\mathsf{y}\}} \mathcal{M}(\pi\varphi_{m,(k,t),\ell};\chi_{m,(k,t),\ell}^{\mathsf{ext}},\kappa_{m,(k,t),\ell}^{\mathsf{ext}}), \label{Deltaxi2varphi}
\end{align}
where 
\begin{align}
    &\chi_{m,(k,t),\ell}^{\mathsf{ext}} = \label{chiextx}  \\
    &\tan^{-1} \left( \frac{\kappa_{m,(k,t),\ell}^{\mathsf{post}}\sin \chi_{m,(k,t),\ell}^{\mathsf{post}} -  \kappa_{m,(k,t),\ell}^{\mathsf{pri}}\sin \chi_{m,(k,t),\ell}^{\mathsf{pri}}}{\kappa_{m,(k,t),\ell}^{\mathsf{post}}\cos \chi_{m,(k,t),\ell}^{\mathsf{post}} -  \kappa_{m,(k,t),\ell}^{\mathsf{pri}}\cos \chi_{m,(k,t),\ell}^{\mathsf{pri}}} \right), \notag \\
    &\kappa_{m,(k,t),\ell}^{\mathsf{ext}} = \notag \\
    &\left[ \left(\kappa_{m,(k,t),\ell}^{\mathsf{post}}\cos \chi_{m,(k,t),\ell}^{\mathsf{post}} -  \kappa_{m,(k,t),\ell}^{\mathsf{pri}}\cos \chi_{m,(k,t),\ell}^{\mathsf{pri}} \right)^2 \right. \label{kappaextx} \\
    & \left. + \left(\kappa_{m,(k,t),\ell}^{\mathsf{post}}\sin \chi_{m,(k,t),\ell}^{\mathsf{post}} -  \kappa_{m,(k,t),\ell}^{\mathsf{pri}}\sin \chi_{m,(k,t),\ell}^{\mathsf{pri}}\right)^2\right]^{\frac{1}{2}}. \notag 
\end{align}
\subsubsection{Messages from $\zeta_{m,(k,t)}$ to $\mathbf{p}_{\mathsf{M};(k,t)}$}
With $\Delta_{\boldsymbol{\varphi}_{m,(k,t)}\rightarrow \zeta_{m,(k,t)}} \allowbreak \! = \! \allowbreak \Delta_{\xi_{m,t}\rightarrow \boldsymbol{\varphi}_{m,(k,t)}}$, we next compute the message from the factor node $\zeta_{m,(k,t)}$ to the variable node $\mathbf{p}_{\mathsf{M};(k,t)}$. Based on \eqref{pVarphipM}, we have 
\begin{align}
    &\Delta_{\zeta_{m,(k,t)}\rightarrow \mathbf{p}_{\mathsf{M};(k,t)}} \notag \\
    &\propto \! \int_{\boldsymbol{\varphi}_{m,(k,t)}} \Delta_{\boldsymbol{\varphi}_{m,(k,t)}\rightarrow \zeta_{m,(k,t)}} p(\boldsymbol{\varphi}_{m,(k,t)}|\mathbf{p}_{\mathsf{M};(k,t)}) \\
    & \propto \prod_{\ell\in \{\mathsf{x},\mathsf{y}\}} \! \mathcal{M} \! \left(\!\frac{\pi\left(\mathbf{p}_{\mathsf{M};(k,t)} \!-\! \mathbf{p}_{\mathsf{B};m}\right)^{\mathsf{T}} \! \boldsymbol{e}_{\ell}}{\|\mathbf{p}_{\mathsf{M};(k,t)} \!-\! \mathbf{p}_{\mathsf{B};m}\|};\chi_{m,(k,t),\ell}^{\mathsf{ext}},\kappa_{m,(k,t),\ell}^{\mathsf{ext}} \!\right)\!. \label{zeta2pM}
\end{align}
Then, the messages $\Delta_{\zeta_{m,(k,t)}\rightarrow \mathbf{p}_{\mathsf{M};(k,t)}}$ are passed into the information fusion module for further processing.

\subsection{Information Fusion Module}
\subsubsection{Messages from $\mathbf{p}_{\mathsf{M};(k,t)}$ to $\eta_{k,t}$}
By collecting the messages $\Delta_{\zeta_{m,(k,t)}\rightarrow \mathbf{p}_{\mathsf{M};(k,t)}}$ at different subarrays, the message from the variable node $\mathbf{p}_{\mathsf{M};(k,t)}$ to the factor node $\eta_{k,t}$ is given by
\begin{equation}\label{DeltapM2eta}
    \Delta_{\mathbf{p}_{\mathsf{M};(k,t)}\rightarrow \eta_{k,t}} \propto \prod_{m\in\mathcal{I}_{M}} \Delta_{\zeta_{m,(k,t)}\rightarrow \mathbf{p}_{\mathsf{M};(k,t)}}.
\end{equation}
Based on \eqref{VMpdf} and \eqref{zeta2pM}, $\Delta_{\mathbf{p}_{\mathsf{M};(k,t)}\rightarrow \eta_{k,t}}$ can be further expressed as 
\begin{align}\label{Deltap2Eta}
    \Delta_{\mathbf{p}_{\mathsf{M};(k,t)}\rightarrow \eta_{k,t}} \propto 
    e^{\varpi_{k,t}(\mathbf{p}_{\mathsf{M};(k,t)})},
\end{align} 
where
\begin{align}
    &\varpi_{k,t}(\mathbf{p}_{\mathsf{M};(k,t)}) \notag \\
    &\triangleq \! \!\sum_{m\in\mathcal{I}_{M}}  \kappa_{m,(k,t),\mathsf{x}}^{\mathsf{ext}} \cos\left(\frac{\pi\left(\mathbf{p}_{\mathsf{M};(k,t)} \!-\! \mathbf{p}_{\mathsf{B};m}\right)^{\mathsf{T}}\boldsymbol{e}_{\mathsf{x}}}{\|\mathbf{p}_{\mathsf{M};(k,t)} \!-\! \mathbf{p}_{\mathsf{B};m}\|} \!-\! \chi_{m,(k,t),\mathsf{x}}^{\mathsf{ext}}\right) \notag \\
    &~ +\! \kappa_{m,(k,t),\mathsf{y}}^{\mathsf{ext}} \! \cos\left(\!\!\frac{\pi\left(\mathbf{p}_{\mathsf{M};(k,t)} \!-\! \mathbf{p}_{\mathsf{B};m}\right)^{\mathsf{T}}\!\boldsymbol{e}_{\mathsf{y}}}{\|\mathbf{p}_{\mathsf{M};(k,t)} \!-\! \mathbf{p}_{\mathsf{B};m}\|} \!-\! \chi_{m,(k,t),\mathsf{y}}^{\mathsf{ext}}\!\!\right)\!. \label{Pikt}
\end{align}
The expression of $\Delta_{\mathbf{p}_{\mathsf{M};(k,t)}\rightarrow \eta_{k,t}}$ in \eqref{Deltap2Eta} is complicated. To simplify the subsequent message calculation, we approximate $\Delta_{\mathbf{p}_{\mathsf{M};(k,t)}\rightarrow \eta_{k,t}}$ by a Gaussian distribution as 
\begin{equation}\label{Deltap2etaGauss}
    \Delta_{\mathbf{p}_{\mathsf{M};(k,t)}\rightarrow \eta_{k,t}}   =  \mathcal{N}  \left(\mathbf{p}_{\mathsf{M};(k,t)};  \boldsymbol{n}_{\mathbf{p}_{\mathsf{M};(k,t)}\rightarrow \eta_{k,t}},  \boldsymbol{C}_{\mathbf{p}_{\mathsf{M};(k,t)}\rightarrow \eta_{k,t}} \right), 
\end{equation}
where $\boldsymbol{n}_{\mathbf{p}_{\mathsf{M};(k,t)}\rightarrow \eta_{k,t}}$ and $\boldsymbol{C}_{\mathbf{p}_{\mathsf{M};(k,t)}\rightarrow \eta_{k,t}}$ are the mean and the covariance matrix, respectively. The values of $\boldsymbol{n}_{\mathbf{p}_{\mathsf{M};(k,t)}\rightarrow \eta_{k,t}}$ and $\boldsymbol{C}_{\mathbf{p}_{\mathsf{M};(k,t)}\rightarrow \eta_{k,t}}$ can be obtained by the Laplace approximation \cite[Chap. 4.4]{bishop2006pattern}. 
{\color{black}Laplace approximation is widely used to find a Gaussian approximation to a continuous pdf, where the mode of the pdf is taken as the mean, and the Hessian matrix at the mode, which reflects the curvature of the pdf, is used to compute the covariance matrix.}
Specifically, $\boldsymbol{n}_{\mathbf{p}_{\mathsf{M};(k,t)}\rightarrow \eta_{k,t}}$ can be derived by finding the maximum of $\varpi_{k,t}(\mathbf{p}_{\mathsf{M};(k,t)})$ over $\mathbf{p}_{\mathsf{M};(k,t)}$, i.e., 
\begin{equation}\label{npM}
    \boldsymbol{n}_{\mathbf{p}_{\mathsf{M};(k,t)}\rightarrow \eta_{k,t}} = \arg \max_{\mathbf{p}_{\mathsf{M};(k,t)}}~ \varpi_{k,t}(\mathbf{p}_{\mathsf{M};(k,t)}).
\end{equation}
Problem \eqref{npM} generally has no closed-form solutions and can be iteratively solved by a general gradient ascent (GA) method. 
For the covariance matrix $\boldsymbol{C}_{\mathbf{p}_{\mathsf{M};(k,t)}\rightarrow \eta_{k,t}}$, we perform the second-order Taylor expansion of $\varpi_{k,t}(\mathbf{p}_{\mathsf{M};(k,t)})$ at $\mathbf{p}_{\mathsf{M};(k,t)} = \boldsymbol{n}_{\mathbf{p}_{\mathsf{M};(k,t)}\rightarrow \eta_{k,t}}$ as 
\begin{align}
    &\varpi_{k,t}(\mathbf{p}_{\mathsf{M};(k,t)}) \notag \\
    &= \varpi_{k,t}(\boldsymbol{n}_{\mathbf{p}_{\mathsf{M};(k,t)}\rightarrow \eta_{k,t}}) + \frac{1}{2} \left(\mathbf{p}_{\mathsf{M};(k,t)}-\boldsymbol{n}_{\mathbf{p}_{\mathsf{M};(k,t)}\rightarrow \eta_{k,t}}\right)^{\mathsf{T}} \notag \\
    &\quad \times \boldsymbol{A}\left(\boldsymbol{n}_{\mathbf{p}_{\mathsf{M};(k,t)}\rightarrow \eta_{k,t}}\right) \left(\mathbf{p}_{\mathsf{M};(k,t)}-\boldsymbol{n}_{\mathbf{p}_{\mathsf{M};(k,t)}\rightarrow \eta_{k,t}}\right),
\end{align}
where $\boldsymbol{A}\left(\boldsymbol{n}_{\mathbf{p}_{\mathsf{M};(k,t)}\rightarrow \eta_{k,t}}\right)$ is the Hessian matrix of $\varpi_{k,t}(\mathbf{p}_{\mathsf{M};(k,t)})$ at $\mathbf{p}_{\mathsf{M};(k,t)} = \boldsymbol{n}_{\mathbf{p}_{\mathsf{M};(k,t)}\rightarrow \eta_{k,t}}$. 
Based on the expression of a Gaussian distribution, the covariance matrix $\boldsymbol{C}_{\mathbf{p}_{\mathsf{M};(k,t)}\rightarrow \eta_{k,t}}$ is given by 
\begin{equation}\label{CpM2eta}
    \boldsymbol{C}_{\mathbf{p}_{\mathsf{M};(k,t)}\rightarrow \eta_{k,t}} = -\boldsymbol{A}\left(\boldsymbol{n}_{\mathbf{p}_{\mathsf{M};(k,t)}\rightarrow \eta_{k,t}}\right)^{-1}.
\end{equation}

\subsubsection{Messages from $\mathbf{p}_{\mathsf{M};k}$ and $\boldsymbol{\theta}_{k}$ to $\eta_{k,t}$}
Based on \eqref{ppMpMkthetak} and the definition of $\mathbf{R}(\boldsymbol{\theta}_k)$ in \eqref{Rdef}, the check function of node $\eta_{k,t}$ is highly nonlinear over $\boldsymbol{\theta}_k$, $k\in \mathcal{I}_K$, $t\in \mathcal{I}_T$. Given $\Delta_{\mathbf{p}_{\mathsf{M};(k,t)}\rightarrow \eta_{k,t}}$, it is difficult to exactly calculate messages $\Delta_{\eta_{k,t}\rightarrow \mathbf{p}_{\mathsf{M};k}}$ and $\Delta_{\eta_{k,t}\rightarrow \boldsymbol{\theta}_{k}}$, as well as the subsequent messages $\Delta_{\mathbf{p}_{\mathsf{M};k} \rightarrow \eta_{k,t}}$ and $\Delta_{\boldsymbol{\theta}_{k} \rightarrow  \eta_{k,t}}$ back to $\eta_{k,t}$, $k\in \mathcal{I}_K$, $t\in \mathcal{I}_T$. 
{\color{black}To simplify the message calculation, we approximate $\Delta_{\mathbf{p}_{\mathsf{M};k} \rightarrow \eta_{k,t}}$ and $\Delta_{\boldsymbol{\theta}_{k} \rightarrow  \eta_{k,t}}$ by Gaussian messages, i.e.,
\begin{align}
    \Delta_{\mathbf{p}_{\mathsf{M};k} \rightarrow \eta_{k,t}} &= \mathcal{N}(\mathbf{p}_{\mathsf{M};k};\boldsymbol{n}_{\mathbf{p}_{\mathbf{p}_{\mathsf{M};k} \rightarrow \eta_{k,t}}},\boldsymbol{C}_{\mathbf{p}_{\mathsf{M};k} \rightarrow \eta_{k,t}}) ~\textrm{and} \label{updpMk} \\
    \Delta_{\boldsymbol{\theta}_{k} \rightarrow  \eta_{k,t}} &= \mathcal{N}(\boldsymbol{\theta}_{k};\boldsymbol{n}_{\boldsymbol{\theta}_{k} \rightarrow  \eta_{k,t}},\boldsymbol{C}_{\boldsymbol{\theta}_{k} \rightarrow  \eta_{k,t}}). \label{updthetak}
\end{align}
The means $\boldsymbol{n}_{\mathbf{p}_{\mathsf{M};k} \rightarrow \eta_{k,t}}$ and $\boldsymbol{n}_{\boldsymbol{\theta}_{k} \rightarrow  \eta_{k,t}}$, as well as the covariance matrices $\boldsymbol{C}_{\mathbf{p}_{\mathsf{M};k} \rightarrow \eta_{k,t}}$ and $\boldsymbol{C}_{\boldsymbol{\theta}_{k} \rightarrow  \eta_{k,t}}$, can be obtained by using the Laplace approximation. Specifically, the means $\boldsymbol{n}_{\mathbf{p}_{\mathsf{M};k} \rightarrow \eta_{k,t}}$ and $\boldsymbol{n}_{\boldsymbol{\theta}_{k} \rightarrow  \eta_{k,t}}$ are given by the MAP estimates of $\mathbf{p}_{\mathsf{M};k}$ and $\boldsymbol{\theta}_{k}$.}
For the message $\Delta_{\mathbf{p}_{\mathsf{M};(k,t)}\rightarrow \eta_{k,t}}$ in \eqref{Deltap2etaGauss}, $\boldsymbol{n}_{\mathbf{p}_{\mathsf{M};(k,t)}\rightarrow \eta_{k,t}}$ is considered as a measurement of $\mathbf{p}_{\mathsf{M};(k,t)}$ corrupted by additive Gaussian noise, with the covariance matrix given by $\boldsymbol{C}_{\mathbf{p}_{\mathsf{M};(k,t)}\rightarrow \eta_{k,t}}$, i.e.,
\begin{equation}
    \boldsymbol{n}_{\mathbf{p}_{\mathsf{M};(k,t)}\rightarrow \eta_{k,t}} = \mathbf{p}_{\mathsf{M};(k,t)} + \boldsymbol{w}_{k,t},
\end{equation}
where $\boldsymbol{w}_{k,t}\sim \mathcal{N}(\boldsymbol{0},\boldsymbol{C}_{\mathbf{p}_{\mathsf{M};(k,t)}\rightarrow \eta_{k,t}})$ denotes the noise vector, $k\in \mathcal{I}_K$, $t\in \mathcal{I}_T$. Since all the MSs share a common transmit pattern, we abbreviate the local position of the $(q(t),s(t))$-th antenna at the $k$-th MS $\mathbf{q}_{\mathsf{M};(k,q(t),s(t))}$ as $\mathbf{q}_{\mathsf{M};t}$. Then, based on \eqref{pMkqs}, the likelihood function of $\mathbf{p}_{\mathsf{M};k}$ and $\boldsymbol{\theta}_{k}$ with the observation $\boldsymbol{n}_{\mathbf{p}_{\mathsf{M};(k,t)}\rightarrow \eta_{k,t}}$ is given by 
\begin{subequations}
    \begin{align}
        & \!\!\!\!\!\! \mathcal{L}(\boldsymbol{n}_{\mathbf{p}_{\mathsf{M};(k,t)}\rightarrow \eta_{k,t}};\mathbf{p}_{\mathsf{M};k},\boldsymbol{\theta}_{k}) \notag \\
        & \!\!\!\!\!\! \triangleq p(\boldsymbol{n}_{\mathbf{p}_{\mathsf{M};(k,t)}\rightarrow \eta_{k,t}}|\mathbf{p}_{\mathsf{M};k},\boldsymbol{\theta}_{k}) \\
        & \!\!\!\!\!\! = \mathcal{N}(\boldsymbol{n}_{\mathbf{p}_{\mathsf{M};(k,t)}\rightarrow \eta_{k,t}}; \! \mathbf{p}_{\mathsf{M};k} \! + \! \mathbf{R}(\boldsymbol{\theta}_k)\mathbf{q}_{\mathsf{M};t}, \! \boldsymbol{C}_{\mathbf{p}_{\mathsf{M};(k,t)}\rightarrow \eta_{k,t}}). \!\!\!
    \end{align}
\end{subequations}
Besides, given $\mathbf{p}_{\mathsf{M};k}$ and $\boldsymbol{\theta}_{k}$, for different $t$, the noisy measurements $\{\boldsymbol{n}_{\mathbf{p}_{\mathsf{M};(k,t)}\rightarrow \eta_{k,t}}\}_{t=1}^T$ are independent of each other. Defined the index set $\mathcal{I}_{T\backslash t} \triangleq \{t'|t'\in \mathcal{I}_T,t'\neq t\}$, and denote by $\mathcal{S}_{{\mathbf{p}_{\mathsf{M};(k,\backslash t)}\rightarrow \eta_{k,\backslash t}}}\triangleq \left\{\boldsymbol{n}_{\mathbf{p}_{\mathsf{M};(k,t')}\rightarrow \eta_{k,t'}}|t'\in \mathcal{I}_{T\backslash t}\right\}$ the set of $\boldsymbol{n}_{\mathbf{p}_{\mathsf{M};(k,t)}\rightarrow \eta_{k,t}}$ except for index $t$, $k\in \mathcal{I}_K$, $t\in \mathcal{I}_T$. The likelihood function of $\mathbf{p}_{\mathsf{M};k}$ and $\boldsymbol{\theta}_{k}$ given the observations $\mathcal{S}_{{\mathbf{p}_{\mathsf{M};(k,\backslash t)}\rightarrow \eta_{k,\backslash t}}}$ is
\begin{multline}\label{bigL}
    \mathcal{L}(\mathcal{S}_{{\mathbf{p}_{\mathsf{M};(k,\backslash t)}\rightarrow \eta_{k,\backslash t}}};\mathbf{p}_{\mathsf{M};k}, \boldsymbol{\theta}_{k}) \\
      \triangleq \prod_{t'\in \mathcal{I}_{T\backslash t}}  \mathcal{L}(\boldsymbol{n}_{\mathbf{p}_{\mathsf{M};(k,t')}\rightarrow \eta_{k,t'}};\mathbf{p}_{\mathsf{M};k}, \boldsymbol{\theta}_{k}).
\end{multline}
Based on the Bayes' theorem, the posterior distribution of $\mathbf{p}_{\mathsf{M};k}$ and $\boldsymbol{\theta}_{k}$ given $\mathcal{S}_{{\mathbf{p}_{\mathsf{M};(k,\backslash t)}\rightarrow \eta_{k,\backslash t}}}$ can be expressed as 
\begin{multline}
    p(\mathbf{p}_{\mathsf{M};k},\boldsymbol{\theta}_{k}|\mathcal{S}_{{\mathbf{p}_{\mathsf{M};(k,\backslash t)}\rightarrow \eta_{k,\backslash t}}})  \\
    \propto \mathcal{L}(\mathcal{S}_{{\mathbf{p}_{\mathsf{M};(k,\backslash t)}\rightarrow \eta_{k,\backslash t}}};\mathbf{p}_{\mathsf{M};k},\boldsymbol{\theta}_{k}) p(\mathbf{p}_{\mathsf{M};k}) p(\boldsymbol{\theta}_k),
\end{multline}
where $p(\mathbf{p}_{\mathsf{M};k})$ and $p(\boldsymbol{\theta}_k)$ are the prior distributions defined in \eqref{ppMk} and \eqref{pthetak}, respectively.
Denote by $\boldsymbol{n}_{\mathbf{p}_{\mathsf{M};k} \rightarrow \eta_{k,t}}$ and $\boldsymbol{n}_{\boldsymbol{\theta}_{k} \rightarrow  \eta_{k,t}}$ the means of $\Delta_{\mathbf{p}_{\mathsf{M};k} \rightarrow \eta_{k,t}}$ and $\Delta_{\boldsymbol{\theta}_{k} \rightarrow  \eta_{k,t}}$, respectively. We approximate $\boldsymbol{n}_{\mathbf{p}_{\mathsf{M};k} \rightarrow \eta_{k,t}}$ and $\boldsymbol{n}_{\boldsymbol{\theta}_{k} \rightarrow \eta_{k,t}}$ by the MAP estimates of $\mathbf{p}_{\mathsf{M};k}$ and $\boldsymbol{\theta}_{k}$ given $\mathcal{S}_{{\mathbf{p}_{\mathsf{M};(k,\backslash t)}\rightarrow \eta_{k,\backslash t}}}$, i.e.,
\begin{equation}\label{MaxJ}
    \left( \! \boldsymbol{n}_{\mathbf{p}_{\mathsf{M};k} \rightarrow \eta_{k,t}}, \boldsymbol{n}_{\boldsymbol{\theta}_k \rightarrow \eta_{k,t}} \! \right) \! = \! \arg \! \max_{\mathbf{p}_{\mathsf{M};k},\boldsymbol{\theta}_{k}}~ \!\! \mathcal{J}(\mathcal{S}_{{\mathbf{p}_{\mathsf{M};(k,\backslash t)}\rightarrow \eta_{k,\backslash t}}};\mathbf{p}_{\mathsf{M};k},\boldsymbol{\theta}_{k}),
\end{equation}
where $\mathcal{J}(\mathcal{S}_{{\mathbf{p}_{\mathsf{M};(k,\backslash t)}\rightarrow \eta_{k,\backslash t}}};\mathbf{p}_{\mathsf{M};k},\boldsymbol{\theta}_{k}) \triangleq \ln (\mathcal{L}(\mathcal{S}_{{\mathbf{p}_{\mathsf{M};(k,\backslash t)}\rightarrow \eta_{k,\backslash t}}};\allowbreak \mathbf{p}_{\mathsf{M};k},\boldsymbol{\theta}_{k}) \allowbreak p(\mathbf{p}_{\mathsf{M};k}) p(\boldsymbol{\theta}_k))$. Problem \eqref{MaxJ} generally has no closed-form solutions, and can be iteratively solved by using GA methods.

{\color{black}
For the covariance matrices $\boldsymbol{C}_{\mathbf{p}_{\mathsf{M};k} \rightarrow \eta_{k,t}}$ and $\boldsymbol{C}_{\boldsymbol{\theta}_{k} \rightarrow  \eta_{k,t}}$, let $\boldsymbol{H}(\boldsymbol{n}_{\mathbf{p}_{\mathsf{M};k} \rightarrow \eta_{k,t}}, \boldsymbol{n}_{\boldsymbol{\theta}_k \rightarrow \eta_{k,t}}) \in \mathbb{R}^{6\times 6}$ denote the Hessian matrix of $\mathcal{J}(\mathcal{S}_{{\mathbf{p}_{\mathsf{M};(k,\backslash t)}\rightarrow \eta_{k,\backslash t}}};\mathbf{p}_{\mathsf{M};k},\boldsymbol{\theta}_{k})$ at $[\boldsymbol{n}_{\mathbf{p}_{\mathsf{M};k} \rightarrow \eta_{k,t}}^{\mathsf{T}}, \boldsymbol{n}_{\boldsymbol{\theta}_k \rightarrow \eta_{k,t}}^{\mathsf{T}}]^{\mathsf{T}}$. By following the Laplace approximation, the covariance matrices $\boldsymbol{C}_{\mathbf{p}_{\mathsf{M};k} \rightarrow \eta_{k,t}}$ and $\boldsymbol{C}_{\boldsymbol{\theta}_{k} \rightarrow  \eta_{k,t}}$ are obtained by 
\begin{equation}
   \!\!\!  \boldsymbol{C}_{\mathbf{p}_{\mathsf{M};k} \rightarrow \eta_{k,t}} \! \triangleq \! \left[\boldsymbol{B}_{k,t}\right]_{1:3,1:3}~  \textrm{and} ~ \boldsymbol{C}_{\boldsymbol{\theta}_{k} \rightarrow  \eta_{k,t}} \! \triangleq \! \left[\boldsymbol{B}_{k,t}\right]_{4:6,4:6},
\end{equation}
respectively, where $\boldsymbol{B}_{k,t} = -\boldsymbol{H}(\boldsymbol{n}_{\mathbf{p}_{\mathsf{M};k} \rightarrow \eta_{k,t}}, \boldsymbol{n}_{\boldsymbol{\theta}_k \rightarrow \eta_{k,t}})^{-1}$.


}


\subsubsection{Messages from $\eta_{k,t}$ to $\mathbf{p}_{\mathsf{M};(k,t)}$}
Based on the sum-product rule, the message from the factor node $\eta_{k,t}$ to the variable node $\mathbf{p}_{\mathsf{M};(k,t)}$ is given by 
\begin{align}\label{etakt2pMkt}
    &\Delta_{\eta_{k,t} \rightarrow \mathbf{p}_{\mathsf{M};(k,t)}} \\
    &\quad\propto \int_{\mathbf{p}_{\mathsf{M};k}} \int_{\boldsymbol{\theta}_{k}}  \Delta_{\mathbf{p}_{\mathsf{M};k} \rightarrow \eta_{k,t}} \Delta_{\boldsymbol{\theta}_{k} \rightarrow  \eta_{k,t}} p\left(\mathbf{p}_{\mathsf{M};(k,t)}|\mathbf{p}_{\mathsf{M};k}, \!\boldsymbol{\theta}_k \right). \notag
\end{align}
From \eqref{ppMpMkthetak}, $p\left(\mathbf{p}_{\mathsf{M};(k,t)}|\mathbf{p}_{\mathsf{M};k}, \!\boldsymbol{\theta}_k \right)$ is a Dirac delta function, and \eqref{etakt2pMkt} is equivalent to compute $p(\mathbf{p}_{\mathsf{M};k} + \mathbf{R}(\boldsymbol{\theta}_k) \mathbf{q}_{\mathsf{M};t})$ conditioned on $\mathbf{p}_{\mathsf{M};k}\sim \Delta_{\mathbf{p}_{\mathsf{M};k} \rightarrow \eta_{k,t}}$ and $\boldsymbol{\theta}_k \sim \Delta_{\boldsymbol{\theta}_{k} \rightarrow  \eta_{k,t}}$. Since $\mathbf{R}(\boldsymbol{\theta}_k)$ is a highly nonlinear function over $\boldsymbol{\theta}_k$, the integral in \eqref{etakt2pMkt} has no closed-form solution. Here, we approximate $\mathbf{R}(\boldsymbol{\theta}_k)$ by the first-order Taylor expansion of $\boldsymbol{\theta}_k$ at $\boldsymbol{n}_{\boldsymbol{\theta}_k \rightarrow \eta_{k,t}}$, i.e.,
\begin{equation}
    \mathbf{R}(\boldsymbol{\theta}_k) = \mathbf{R}(\boldsymbol{n}_{\boldsymbol{\theta}_k \rightarrow \eta_{k,t}}) + \!\!\!\! \sum_{\ell\in \{\mathsf{x},\mathsf{y},\mathsf{z}\}} \!\!\!\! \mathbf{R}'_{\ell}(\boldsymbol{n}_{\boldsymbol{\theta}_k \rightarrow \eta_{k,t}}) (\theta_{\ell;k} - n_{\theta_{\ell;k} \rightarrow \eta_{k,t}} ) ,
\end{equation}
where $\mathbf{R}'_{\ell}(\boldsymbol{n}_{\boldsymbol{\theta}_k \rightarrow \eta_{k,t}}) = \frac{\partial \mathbf{R}(\boldsymbol{\theta}_k)}{\partial \theta_{\ell;k}} \big|_{\boldsymbol{\theta}_k = \boldsymbol{n}_{\boldsymbol{\theta}_k \rightarrow \eta_{k,t}}}$, $\ell\in \{\mathsf{x},\mathsf{y},\mathsf{z}\}$.
Then, $\mathbf{p}_{\mathsf{M};k} + \mathbf{R}(\boldsymbol{\theta}_k) \mathbf{q}_{\mathsf{M};t}$ is reduced to a linear function over $\mathbf{p}_{\mathsf{M};k}$ and $\boldsymbol{\theta}_k$. Since $\Delta_{\mathbf{p}_{\mathsf{M};k} \rightarrow \eta_{k,t}}$ and $\Delta_{\boldsymbol{\theta}_{k} \rightarrow  \eta_{k,t}}$ are Gaussian messages, $\mathbf{p}_{\mathsf{M};k} + \mathbf{R}(\boldsymbol{\theta}_k) \mathbf{q}_{\mathsf{M};t}$ also follows a Gaussian distribution with the mean given by 
\begin{equation}
    \boldsymbol{n}_{\eta_{k,t} \rightarrow \mathbf{p}_{\mathsf{M};(k,t)}} = \boldsymbol{n}_{\mathbf{p}_{\mathsf{M};k} \rightarrow \eta_{k,t}} + \mathbf{R}(\boldsymbol{n}_{\boldsymbol{\theta}_{k} \rightarrow \eta_{k,t}})\mathbf{q}_{\mathsf{M};t},
\end{equation}
and the covariance matrix given by 
\begin{equation}\label{Cetakt}
    \boldsymbol{C}_{\eta_{k,t} \rightarrow \mathbf{p}_{\mathsf{M};(k,t)}}  =  \boldsymbol{C}_{\mathbf{p}_{\mathsf{M};k} \rightarrow \eta_{k,t}} +  \boldsymbol{Q}_{k,t} \boldsymbol{C}_{\boldsymbol{\theta}_{k} \rightarrow  \eta_{k,t}} \boldsymbol{Q}_{k,t}^{\mathsf{T}},
\end{equation}
where $\boldsymbol{Q}_{k,t} \triangleq [ \mathbf{R}'_{\mathsf{x}}(\boldsymbol{n}_{\boldsymbol{\theta}_k \rightarrow \eta_{k,t}}) \mathbf{q}_{\mathsf{M};t}, \allowbreak \mathbf{R}'_{\mathsf{y}}(\boldsymbol{n}_{\boldsymbol{\theta}_k \rightarrow \eta_{k,t}}) \mathbf{q}_{\mathsf{M};t}, \allowbreak \mathbf{R}'_{\mathsf{z}}(\boldsymbol{n}_{\boldsymbol{\theta}_k \rightarrow \eta_{k,t}}) \mathbf{q}_{\mathsf{M};t}  ]$.
Therefore, $\Delta_{\eta_{k,t} \rightarrow \mathbf{p}_{\mathsf{M};(k,t)}}$ is approximated by a Gaussian distribution as 
\begin{equation}\label{Deltaeta2pMGauss}
    \Delta_{\eta_{k,t} \rightarrow \mathbf{p}_{\mathsf{M};(k,t)}}  =  \mathcal{N} \left(\mathbf{p}_{\mathsf{M};(k,t)};\boldsymbol{n}_{\eta_{k,t} \rightarrow \mathbf{p}_{\mathsf{M};(k,t)}},\boldsymbol{C}_{\eta_{k,t} \rightarrow \mathbf{p}_{\mathsf{M};(k,t)}} \right).  
\end{equation}

\subsubsection{Messages from $\mathbf{p}_{\mathsf{M};(k,t)}$ to $\zeta_{m,(k,t)}$}
We next consider the messages back to the AoA estimation module. Define the index sets $\mathcal{I}_{M\backslash m}=\left\{ m'|m'\in\mathcal{I}_{M}, m'\neq m\right\}$. Based on the sum-product rule, the message from the variable node $\mathbf{p}_{\mathsf{M};(k,t)}$ to the factor node $\zeta_{m,(k,t)}$ is given by 
\begin{align}\label{DeltapM2zeta}
    \Delta_{\mathbf{p}_{\mathsf{M};(k,t)} \rightarrow \zeta_{m,(k,t)}} &\propto \Delta_{\eta_{k,t} \rightarrow \mathbf{p}_{\mathsf{M};(k,t)}} \prod_{m' \in \mathcal{I}_{M\backslash m}} \Delta_{\zeta_{m',(k,t)}\rightarrow \mathbf{p}_{\mathsf{M};(k,t)}} \notag \\
    &= \Delta_{\eta_{k,t} \rightarrow \mathbf{p}_{\mathsf{M};(k,t)}} \Gamma_{m,(k,t)} ,
\end{align}
where $\Gamma_{m,(k,t)} \triangleq \prod_{m' \in \mathcal{I}_{M\backslash m}} \Delta_{\zeta_{m',(k,t)}\rightarrow \mathbf{p}_{\mathsf{M};(k,t)}}$. Note that $\Gamma_{m,(k,t)}$ has a similar form to the RHS of \eqref{DeltapM2eta}. To simplify the message calculation, we approximate $\Gamma_{m,(k,t)}$ by a Gaussian distribution as
\begin{equation}\label{Gammamkt}
    \Gamma_{m,(k,t)} = \mathcal{N}(\mathbf{p}_{\mathsf{M};(k,t)};\bar{\Gamma}_{m,(k,t)},\boldsymbol{\Sigma}_{m,(k,t)}),
\end{equation}
where the mean $\bar{\Gamma}_{m,(k,t)}$ and the covariance matrix $\boldsymbol{\Sigma}_{m,(k,t)}$ can be obtained by a similar manner as shown in \eqref{npM} and \eqref{CpM2eta}. Based on \eqref{Deltaeta2pMGauss}, \eqref{DeltapM2zeta}, \eqref{Gammamkt}, and the closure of Gaussian distributions under multiplication \cite[Chap. 8.1.8]{petersenMatrixCookbook}, $\Delta_{\mathbf{p}_{\mathsf{M};(k,t)} \rightarrow \zeta_{m,(k,t)}}$ is also a Gaussian distribution as expressed in \eqref{Deltapm2zeta}, with the mean $\boldsymbol{n}_{\mathbf{p}_{\mathsf{M};(k,t)} \rightarrow \zeta_{m,(k,t)}}$ and the covariance matrix $\boldsymbol{C}_{\mathbf{p}_{\mathsf{M};(k,t)} \rightarrow \zeta_{m,(k,t)}}$ given by 
\begin{align}
    &\boldsymbol{n}_{\mathbf{p}_{\mathsf{M};(k,t)} \rightarrow \zeta_{m,(k,t)}} = \boldsymbol{C}_{\mathbf{p}_{\mathsf{M};(k,t)} \rightarrow \zeta_{m,(k,t)}} \boldsymbol{C}_{\eta_{k,t} \rightarrow \mathbf{p}_{\mathsf{M};(k,t)}}^{-1} \notag \\
    &~~  \times \boldsymbol{n}_{\eta_{k,t} \rightarrow \mathbf{p}_{\mathsf{M};(k,t)}} + \boldsymbol{C}_{\mathbf{p}_{\mathsf{M};(k,t)} \rightarrow \zeta_{m,(k,t)}} \boldsymbol{\Sigma}_{m,(k,t)}^{-1} \bar{\Gamma}_{m,(k,t)} \label{npM2zeta}\\
    &\boldsymbol{C}_{\mathbf{p}_{\mathsf{M};(k,t)} \rightarrow \zeta_{m,(k,t)}} = \left(\boldsymbol{C}_{\eta_{k,t} \rightarrow \mathbf{p}_{\mathsf{M};(k,t)}}^{-1} + \boldsymbol{\Sigma}_{m,(k,t)}^{-1} \right)^{-1}. \label{CpM2zeta}
\end{align}

\subsection{Final Estimation Output}
After reaching the maximum number of iterations between the AoA estimation module and the information fusion module, the algorithm outputs the final estimate of $\mathbf{p}_{\mathsf{M};k}$ and $\boldsymbol{\theta}_{k}$, $k\in \mathcal{I}_{K}$. Define $\mathcal{S}_{{\mathbf{p}_{\mathsf{M};k}\rightarrow \eta_{k}}}\triangleq \left\{\boldsymbol{n}_{\mathbf{p}_{\mathsf{M};(k,t)}\rightarrow \eta_{k,t}}|t\in \mathcal{I}_{T}\right\}$ the set of $\boldsymbol{n}_{\mathbf{p}_{\mathsf{M};(k,t)}\rightarrow \eta_{k,t}}$ for all $t\in \mathcal{I}_T$, $k\in \mathcal{I}_K$. Similar as in \eqref{MaxJ}, the final estimate $\hat{\mathbf{p}}_{\mathsf{M};k}$ and $\hat{\boldsymbol{\theta}}_{k}$ are obtained based on the MAP principle by solving
\begin{align}\label{MAPfinal}
    \left(\hat{\mathbf{p}}_{\mathsf{M};k}, \hat{\boldsymbol{\theta}}_{k} \right)&= \arg \max_{\mathbf{p}_{\mathsf{M};k},\boldsymbol{\theta}_{k}}~ \mathcal{Q}(\mathcal{S}_{{\mathbf{p}_{\mathsf{M};(k, t)}\rightarrow \eta_{k, t}}};\mathbf{p}_{\mathsf{M};k},\boldsymbol{\theta}_{k}),
\end{align}
where
\begin{align}
    &\mathcal{Q}(\mathcal{S}_{{\mathbf{p}_{\mathsf{M};(k, t)}\rightarrow \eta_{k, t}}};\mathbf{p}_{\mathsf{M};k},\boldsymbol{\theta}_{k})  \\
    &= \!\sum_{t\in \mathcal{I}_{T}} \ln \mathcal{L}(\boldsymbol{n}_{\mathbf{p}_{\mathsf{M};(k,t)}\rightarrow \eta_{k,t}};\mathbf{p}_{\mathsf{M};k},\boldsymbol{\theta}_{k}) \! + \! \ln p(\mathbf{p}_{\mathsf{M};k}) \! + \! \ln p(\boldsymbol{\theta}_k). \notag
\end{align}
Problem \eqref{MAPfinal} can be solved by using GA methods.

\subsection{Overall Algorithm}

\begin{algorithm}[t]
    \small{
    \linespread{1.2}\selectfont
    \LinesNumbered
    \KwIn{The received signal $\boldsymbol{Y}$, the transmit pattern $\boldsymbol{L}$.}
    \KwIni{Initialize $\Delta_{\mathbf{p}_{\mathsf{M};(k,t)} \rightarrow \zeta_{m,(k,t)}}  = \mathcal{N}(\mathbf{p}_{\mathsf{M};(k,t)}; \allowbreak \boldsymbol{n}_{\mathbf{p}_{\mathsf{M};(k,t)} \rightarrow \zeta_{m,(k,t)}}, \allowbreak \boldsymbol{C}_{\mathbf{p}_{\mathsf{M};(k,t)} \rightarrow \zeta_{m,(k,t)}})$, $\forall m\in \mathcal{I}_M$, $\forall k\in \mathcal{I}_K$, $\forall t\in \mathcal{I}_T$}
    \While{the stopping criterion is not met}{
    \textbf{\% AoA Estimation Module\;}
    \% $\forall m\in \mathcal{I}_M$, $\forall k\in \mathcal{I}_K$, $\forall t\in \mathcal{I}_T$\;
    Update the message $\Delta_{\zeta_{m,(k,t)} \rightarrow \boldsymbol{\varphi}_{m,(k,t)}}$ by \eqref{Gauss2VM}\;
    Compute the posterior distribution $p(\boldsymbol{\varphi}_{m,(k,t)}|\boldsymbol{Y}_{m,t})$ by calling the VALSE algorithm\;
    Update the message $\Delta_{\xi_{m,t}\rightarrow \boldsymbol{\varphi}_{m,(k,t)}}$ by \eqref{Deltaxi2varphi}\;
    Update the message $\Delta_{\zeta_{m,(k,t)}\rightarrow \mathbf{p}_{\mathsf{M};(k,t)}}$ by \eqref{zeta2pM}\;
    \textbf{\% Information Fusion Module\;}
    \% $\forall k\in \mathcal{I}_K$, $\forall t\in \mathcal{I}_T$\;
    Update the message $\Delta_{\mathbf{p}_{\mathsf{M};(k,t)}\rightarrow \eta_{k,t}}$ by \eqref{Deltap2etaGauss}\;
    Update the messages $\Delta_{\mathbf{p}_{\mathsf{M};k} \rightarrow \eta_{k,t}}$ and $\Delta_{\boldsymbol{\theta}_k \rightarrow \eta_{k,t}}$ by \eqref{updpMk} and \eqref{updthetak}\;
    Update the message $\Delta_{\eta_{k,t} \rightarrow \mathbf{p}_{\mathsf{M};(k,t)}}$ by \eqref{Deltaeta2pMGauss}\;
    Update the message $\Delta_{\mathbf{p}_{\mathsf{M};(k,t)} \rightarrow \zeta_{m,(k,t)}}$ by \eqref{Deltapm2zeta}\;
    }
    Solve \eqref{MAPfinal} for the final estimate $\hat{\mathbf{p}}_{\mathsf{M};k}$ and $\hat{\boldsymbol{\theta}}_{k}$, $\forall k\in \mathcal{I}_K$\;
    \KwOut{$\hat{\mathbf{p}}_{\mathsf{M};k}$ and $\hat{\boldsymbol{\theta}}_{k}$, or $\hat{\mathbf{p}}_{\mathsf{M};k}$ and $\mathbf{R}(\hat{\boldsymbol{\theta}}_{k})$, $\forall k\in \mathcal{I}_K$}
    \caption{The proposed APPLE algorithm}
    \label{OverallAlgo}
    }
\end{algorithm}

Based on the discussions in the preceding subsections, the proposed APPLE algorithm is summarized in Algorithm \ref{OverallAlgo}. The input of the APPLE algorithm includes the received signals at the BS $\boldsymbol{Y}$, and the predetermined transmit pattern $\boldsymbol{L}$. The APPLE algorithm finally returns the estimate of $K$ MSs' positions and attitudes. The APPLE algorithm starts with the initialization of $\Delta_{\mathbf{p}_{\mathsf{M};(k,t)} \rightarrow \zeta_{m,(k,t)}}$ to Gaussian distributions, $\forall k\in \mathcal{I}_K$, $\forall t\in \mathcal{I}_T$. When there is no strong prior knowledge about the MSs' positions and attitudes, the mean vector $\boldsymbol{n}_{\mathbf{p}_{\mathsf{M};(k,t)} \rightarrow \zeta_{m,(k,t)}}$ and the covariance matrix $\boldsymbol{C}_{\mathbf{p}_{\mathsf{M};(k,t)} \rightarrow \zeta_{m,(k,t)}}$ can be initialized to $[0,0,1]^{\mathsf{T}}$ and $\sigma_{\mathsf{ini}}^2 \boldsymbol{I}$, respectively, where $\sigma_{\mathsf{ini}}$ is a large positive value, $\forall m\in \mathcal{I}_M$, $\forall k\in \mathcal{I}_K$, $\forall t\in \mathcal{I}_T$. The APPLE algorithm executed as the iterative message passing between the AoA estimation module and the information fusion module. Specifically, in Algorithm \ref{OverallAlgo}, line 2 to line 7 correspond to the calculation of message passing in the AoA estimation module, and line 8 to line 13 correspond to that in the information fusion module. The stopping criterion is generally set as the maximum number of iteration being reached. Then, the APPLE algorithm solves \eqref{MAPfinal} for the final estimate of the MSs' positions and attitudes, and output $\hat{\mathbf{p}}_{\mathsf{M};k}$ and $\hat{\boldsymbol{\theta}}_{k}$, or equivalently, $\hat{\mathbf{p}}_{\mathsf{M};k}$ and $\mathbf{R}(\hat{\boldsymbol{\theta}}_{k})$, $\forall k\in \mathcal{I}_K$.



\section{Estimation-Error Lower Bound}\label{SecLB}
In this section, we derive an estimation-error lower bound for evaluating the PAE performance of the proposed APPLE algorithm. We assume that the prior distributions assigned to MSs' positions and attitudes in \eqref{ppMk} and \eqref{pthetak} are non-informative, and the error bound is derived by following the idea of misspecified Cram\'er Rao bound (MCRB) \cite{fortunatiPerformanceBoundsParameter2017,chenChannelModelMismatch2022}. 
{\color{black} The MCRB is an extension of the conventional CRB, used to evaluate the parameter estimation performance when model assumptions are incorrect or misspecified. Specifically, if there exists a unique parameter vector in the misspecified parameter space that minimizes the Kullback-Leibler (KL) divergence between the likelihood function based on the misspecified model and the true distribution of observations, this parameter vector is referred to as the \textit{pseudotrue parameter vector}. The MCRB provides a lower bound on the MSE of any estimator obtained from the misspecified model with respect to this pseudotrue parameter vector. Moreover, a lower bound on the MSE of any estimator from the misspecified model with respect to the true model parameters can be obtained \cite{fortunatiPerformanceBoundsParameter2017}.}
In this work, denote $\boldsymbol{\gamma} \triangleq \left[\mathbf{p}_{\mathsf{M};1}^{\mathsf{T}}, \ldots, \mathbf{p}_{\mathsf{M};K}^{\mathsf{T}}, \boldsymbol{\theta}_1^{\mathsf{T}},\ldots,\boldsymbol{\theta}_K^{\mathsf{T}}\right]^{\mathsf{T}} \in \mathbb{R}^{6K}$
as the position and attitude parameter vector of all MSs.
Based on the near-field received signal model \eqref{Y}, 
the likelihood function is given by
\begin{equation}
    p(\boldsymbol{Y};\boldsymbol{\gamma}) = \prod_{t\in \mathcal{I}_T} \mathcal{CN}\Big(\boldsymbol{y}_t - \sum_{k\in \mathcal{I}_K} \boldsymbol{h}_{(k,t)} x_{k,t}; \boldsymbol{0}, \sigma_w^2 \boldsymbol{I} \Big).
\end{equation}
Similarly, define a vector 
\begin{equation}
    \boldsymbol{\gamma}_{\mathsf{FF}} \triangleq \left[\boldsymbol{\gamma}, \varrho_{1,(1,1)}^{\mathsf{Re}},\varrho_{1,(1,1)}^{\mathsf{Im}},\ldots, \varrho_{M,(K,T)}^{\mathsf{Re}},\varrho_{M,(K,T)}^{\mathsf{Im}}\right]^{\mathsf{T}},
\end{equation}
which contains the real parameters to be estimated under the SWFF signal model \eqref{SubFarSigMod}. Based on the SWFF received signal model \eqref{SubFarSigMod}, the likelihood function is given by 
\begin{align}
    p(\boldsymbol{Y};\boldsymbol{\gamma}_{\mathsf{FF}}) &= \prod_{m\in \mathcal{I}_{M}} \prod_{t\in \mathcal{I}_{T}} \mathcal{CN}\Big(\mathrm{vec}\Big(\boldsymbol{Y}_{m,t} \notag \\
    &\quad - \sum_{k\in\mathcal{I}_K} \varrho_{m,(k,t)}\boldsymbol{\varUpsilon}_{m,(k,t)}\Big); \boldsymbol{0},\sigma_w^2 \boldsymbol{I}\Big).
\end{align}
Denote by $\tilde{\boldsymbol{\gamma}}$ and $\tilde{\boldsymbol{\gamma}}_{\mathsf{FF}}$ the ground truth values of $\boldsymbol{\gamma}$ and $\boldsymbol{\gamma}_{\mathsf{FF}}$, respectively. Under the SWFF signal model, an estimation error lower bound over $\boldsymbol{\gamma}$ can be given by \cite{chenChannelModelMismatch2022}
\begin{multline}
    \textrm{LB}(\tilde{\boldsymbol{\gamma}},\boldsymbol{\gamma}_{\mathsf{FF},0}) = \boldsymbol{A}_{\boldsymbol{\gamma}_{\mathsf{FF},0}}^{-1} \boldsymbol{B}_{\boldsymbol{\gamma}_{\mathsf{FF},0}}^{-1} \boldsymbol{A}_{\boldsymbol{\gamma}_{\mathsf{FF},0}}^{-1} \\
     + (\boldsymbol{\gamma}_{\mathsf{FF},0}-\tilde{\boldsymbol{\gamma}}_{\mathsf{FF}})(\boldsymbol{\gamma}_{\mathsf{FF},0}-\tilde{\boldsymbol{\gamma}}_{\mathsf{FF}})^{\mathsf{T}},
\end{multline}
where $\boldsymbol{\gamma}_{\mathsf{FF},0}$ is the pseudotrue parameter vector of the SWFF signal model that minimizes the KL divergence between $p(\boldsymbol{Y};\boldsymbol{\gamma}_{\mathsf{FF}})$ and $p(\boldsymbol{Y};\tilde{\boldsymbol{\gamma}})$ \cite{fortunatiPerformanceBoundsParameter2017}. Based on \cite{chenChannelModelMismatch2022}, 
this KL divergence minimization problem is equivalent to 
\begin{equation}\label{minikld}
    \boldsymbol{\gamma}_{\mathsf{FF},0} = \arg\min_{\boldsymbol{\gamma}_{\mathsf{FF}}}~ \sum_{t\in\mathcal{I}_{T}}\|\boldsymbol{\Omega}_{t}(\tilde{\boldsymbol{\gamma}}) - \boldsymbol{\Omega}_{\mathsf{FF},t}(\boldsymbol{\gamma}_{\mathsf{FF}})\|_{\mathsf{F}},
\end{equation}
where the $(u,v)$-th element of $\boldsymbol{\Omega}_{t}(\tilde{\boldsymbol{\gamma}})$ is given by $\sum_{k\in \mathcal{I}_{K}} h_{(u,v),(k,t)} x_{k,t} \Big|_{\boldsymbol{\gamma} = \tilde{\boldsymbol{\gamma}}}$, and the $(u,v)$-th element of $\boldsymbol{\Omega}_{\mathsf{FF},t}(\boldsymbol{\gamma}_{\mathsf{FF}})$ is given by the $(m,\imath,\jmath)$-th element of $\sum_{k\in\mathcal{I}_K} \varrho_{m,(k,t)} \boldsymbol{\varUpsilon}_{m,(k,t)}\left(\boldsymbol{\varphi}_{m,(k,t)}\right)$ with $(m,\imath,\jmath)=\mathcal{G}(u,v)$, $u\in \mathcal{I}_{N_{\mathsf{B};\mathsf{x}}}$, $v\in\mathcal{I}_{N_{\mathsf{B};\mathsf{y}}}$, $m\in \mathcal{I}_M$, $\imath \in \mathcal{I}_{N_{m,\mathsf{x}}}$, $\jmath \in \mathcal{I}_{N_{m,\mathsf{y}}}$, $k\in \mathcal{I}_{K}$. 
Problem \eqref{minikld} generally lacks a closed-form solution and can be solved numerically by exhaustive search.
$\boldsymbol{A}_{\boldsymbol{\gamma}_{\mathsf{FF},0}}$ and $\boldsymbol{B}_{\boldsymbol{\gamma}_{\mathsf{FF},0}}$ are two generalizations of the Fisher information matrix under the SWFF signal model. {\color{black}Let $\boldsymbol{\mu}(\boldsymbol{\gamma}) \triangleq \left[ \textrm{vec}(\boldsymbol{\Omega}_{1}(\boldsymbol{\gamma}))^{\mathsf{T}}, \ldots, \textrm{vec}(\boldsymbol{\Omega}_{T}(\boldsymbol{\gamma}))^{\mathsf{T}}\right]^{\mathsf{T}}$ and $\boldsymbol{\mu}_{\mathsf{FF}}(\boldsymbol{\gamma}_{\mathsf{FF}}) \allowbreak \triangleq \allowbreak \left[ \textrm{vec}(\boldsymbol{\Omega}_{\mathsf{FF},1}(\boldsymbol{\gamma}_{\mathsf{FF}}))^{\mathsf{T}}, \ldots, \textrm{vec}(\boldsymbol{\Omega}_{\mathsf{FF},T}(\boldsymbol{\gamma}_{\mathsf{FF}}))^{\mathsf{T}}\right]^{\mathsf{T}}$.}
$\boldsymbol{A}_{\boldsymbol{\gamma}_{\mathsf{FF},0}}$ and $\boldsymbol{B}_{\boldsymbol{\gamma}_{\mathsf{FF},0}}$ are given by 
\begin{align}
    \left[\boldsymbol{A}_{\boldsymbol{\gamma}_{\mathsf{FF},0}}\right]_{a,b} &\triangleq \frac{2}{\sigma_w^2} \textrm{Re}\left\{ \boldsymbol{\epsilon}^{\mathsf{H}}(\boldsymbol{\gamma}_{\mathsf{FF}}) \frac{\partial^2 \boldsymbol{\mu}_{\mathsf{FF}}(\boldsymbol{\gamma}_{\mathsf{FF}})}{\partial \left[\boldsymbol{\gamma}_{\mathsf{FF}}\right]_{a} \partial \left[\boldsymbol{\gamma}_{\mathsf{FF}}\right]_{b}}  \right. \notag \\
    &\quad - \left.  \frac{\partial \boldsymbol{\mu}_{\mathsf{FF}}^{\mathsf{H}}(\boldsymbol{\gamma}_{\mathsf{FF}})}{\partial \left[\boldsymbol{\gamma}_{\mathsf{FF}}\right]_{a}} \frac{\partial \boldsymbol{\mu}_{\mathsf{FF}}(\boldsymbol{\gamma}_{\mathsf{FF}})}{\partial \left[\boldsymbol{\gamma}_{\mathsf{FF}}\right]_{b}} \right\} \bigg|_{\boldsymbol{\gamma}_{\mathsf{FF}} = \boldsymbol{\gamma}_{\mathsf{FF},0}}, \\
    \left[\boldsymbol{B}_{\boldsymbol{\gamma}_{\mathsf{FF},0}}\right]_{a,b} &\triangleq \frac{4}{\sigma_w^2} \textrm{Re}\left\{\boldsymbol{\epsilon}^{\mathsf{H}}(\boldsymbol{\gamma}_{\mathsf{FF}}) \frac{\partial \boldsymbol{\mu}_{\mathsf{FF}}(\boldsymbol{\gamma}_{\mathsf{FF}})}{\partial \left[\boldsymbol{\gamma}_{\mathsf{FF}}\right]_{a}}\right\} \notag \\
    &\quad \textrm{Re}\left\{\boldsymbol{\epsilon}^{\mathsf{H}}(\boldsymbol{\gamma}_{\mathsf{FF}}) \frac{\partial \boldsymbol{\mu}_{\mathsf{FF}}(\boldsymbol{\gamma}_{\mathsf{FF}})}{\partial \left[\boldsymbol{\gamma}_{\mathsf{FF}}\right]_{b}}\right\} \bigg|_{\boldsymbol{\gamma}_{\mathsf{FF}} = \boldsymbol{\gamma}_{\mathsf{FF},0}}  \\
    &\quad + \frac{2}{\sigma_w^2} \textrm{Re}\left\{\frac{\partial \boldsymbol{\mu}_{\mathsf{FF}}^{\mathsf{H}}(\boldsymbol{\gamma}_{\mathsf{FF}})}{\partial \left[\boldsymbol{\gamma}_{\mathsf{FF}}\right]_{a}} \frac{\partial \boldsymbol{\mu}_{\mathsf{FF}}(\boldsymbol{\gamma}_{\mathsf{FF}})}{\partial \left[\boldsymbol{\gamma}_{\mathsf{FF}}\right]_{b}}\right\} \bigg|_{\boldsymbol{\gamma}_{\mathsf{FF}} = \boldsymbol{\gamma}_{\mathsf{FF},0}}, \notag
\end{align}
where $\boldsymbol{\epsilon}(\boldsymbol{\gamma}_{\mathsf{FF}}) \triangleq \boldsymbol{\mu}(\tilde{\boldsymbol{\gamma}}) - \boldsymbol{\mu}_{\mathsf{FF}}(\boldsymbol{\gamma}_{\mathsf{FF}})$, $a, b\in \mathcal{I}_{6K+2MKT}$. Denote by $\hat{\boldsymbol{\gamma}}$ an estimate of $\boldsymbol{\gamma}$ obtained under the SWFF signal model, the MSE of $\hat{\boldsymbol{\gamma}}$ over $\boldsymbol{\gamma}$ is bounded by the LB, given by
\begin{equation}\label{LBexpression}
    \mathbb{E}\left[ (\hat{\boldsymbol{\gamma}}-\tilde{\boldsymbol{\gamma}})(\hat{\boldsymbol{\gamma}}-\tilde{\boldsymbol{\gamma}})^{\mathsf{T}} \right] \succeq \left[\textrm{LB}(\tilde{\boldsymbol{\gamma}},\boldsymbol{\gamma}_{\mathsf{FF},0})\right]_{1:6K,1:6K}.
\end{equation}
The obtained LB can be used to evaluate the PAE performance of the proposed APPLE algorithm.

\section{Numerical Results}

\subsection{System Parameters and Performance Metric}

We evaluate the performance of the proposed APPLE algorithm under varying parameter configurations.
We assume that both the BS array and the MS array are of square shape with $N_{\mathsf{B};\mathsf{x}} = N_{\mathsf{B};\mathsf{y}} = N_{\mathsf{B}}$ and $N_{\mathsf{M};\mathsf{x}} = N_{\mathsf{M};\mathsf{y}} = N_{\mathsf{M}}$. 
Unless otherwise specified, the index set of the activated antenna at each MS array, i.e., the transmit pattern, is set to $\mathcal{T}_5=\{(1,1),(1,N_{\mathsf{M}}),(N_{\mathsf{M}},1),(N_{\mathsf{M}},N_{\mathsf{M}}),\allowbreak (\lceil \frac{N_{\mathsf{M}}+1}{2},\frac{N_{\mathsf{M}}+1}{2} \rceil)\}$ by default. The azimuth and elevation angles of the BS-MS direction, as annotated in Fig. \ref{scene}, are randomly drawn from $\alpha_{\mathsf{azi};k}\in [0,2\pi)$ and $\alpha_{\mathsf{ele};k}\in [\pi/12,\pi/2]$, $k\in \mathcal{I}_K$. 
The MS array rotation angles are randomly drawn from $\theta_{\mathsf{x};k}\in [-\pi,\pi)$, $\theta_{\mathsf{y};k}\in [-\pi/3,\pi/3]$, and $\theta_{\mathsf{z};k}\in [-\pi,\pi)$, $k\in \mathcal{I}_K$. The carrier frequency is set to $f$=$28\textrm{GHz}$. The AWGN noise power is set to $\sigma_w^2$=$-70\textrm{dBm}$. The received signals are generated based on the near-field signal model in \eqref{Y}.
The channel coefficient is generated by \eqref{hcoeff}.
For simplicity, the antenna gain $\beta_k$ is set to unit $1$ for all $k\in \mathcal{I}_K$. The signals transmitted by all activated MS antennas are set to $x_{k,t}=\sqrt{P_x}$ for all $k \in \mathcal{I}_K$ and $t \in \mathcal{I}_T$, where $P_x$ is the transmit power.

The performance of the proposed APPLE algorithm is measured by the root-mean-square-error (RMSE) of the estimates of $\mathbf{p}_{\mathsf{M}}$ and the normalized mean-square-error (NMSE) of the estimates of $\mathbf{R}$, respectively defined as 
\begin{align}
    \textrm{RMSE}(\mathbf{p}_{\mathsf{M}}) &\triangleq \sqrt{\mathbb{E}\left[ \sum_{k\in \mathcal{I}_K} \| \mathbf{p}_{\mathsf{M};k} - \hat{\mathbf{p}}_{\mathsf{M};k} \|^2  \right]}~ \textrm{and} \label{RMSEpm} \\
    \textrm{NMSE}(\mathbf{R}) &\triangleq \mathbb{E}\left[ \sum_{k\in \mathcal{I}_K} \frac{\| \mathbf{R}(\boldsymbol{\theta}_k) - \mathbf{R}(\hat{\boldsymbol{\theta}}_k) \|_{\textrm{F}}^2}{\| \mathbf{R}(\boldsymbol{\theta}_k) \|_{\textrm{F}}^2}  \right]. \label{RMSER}
\end{align}
The expectations in \eqref{RMSER} are numerically approximated by averaging the results of $500$ random experiments. 


{\color{black}
When selecting the subarray sizes $N_{m,\mathsf{x}}$ and $N_{m,\mathsf{y}}$, Assumption \ref{SubArrAssump} suggests that these dimensions should be sufficiently small to ensure that the MS in the far-field region of each subarray. Meanwhile, as illustrated later in Fig. \ref{verPartNum}, overly small subarrays can introduce excessive nuisance parameters, thereby degrading APPLE's performance. In practical deployments, a conservative partitioning strategy can be adopted based on the minimum BS-MS distance. Typically, BS arrays are deployed at elevated positions to mitigate obstructions and extend coverage. For example, standard BS heights in 5G scenarios are approximately 10m for urban micro cells, 25m for urban macro cells, and 2–3 m for indoor deployments \cite{TR38901}. This setup establishes a minimum BS-MS distance $r_{\mathsf{min}}$, assuming the MSs are positioned at ground level. 
For instance, consider an indoor scenario where the BS height is 3 m and the MS height is 1.5 m, the minimum BS-MS distance is $r_{\mathsf{min}}=1.5$m. At a carrier frequency of $f=28$GHz, ensuring that all potential MSs are in the far-field region of each BS subarray requires that the largest dimension of each subarray, $S_{m}$, satisfies $D_{\mathsf{R},m} = \frac{2S_m^2}{\lambda} \leq r_{\mathsf{min}},~ \forall m\in \mathcal{I}_{M}$. This implies that $S_m \allowbreak \leq \allowbreak \sqrt{\frac{\lambda r_{\mathsf{min}}}{2}}=0.09$m. Meanwhile, $S_m$ should not be significantly smaller than $\sqrt{\frac{\lambda r_{\mathsf{min}}}{2}}$ to avoid performance degradation. To balance these considerations, the BS array is always uniformly partitioned into an appropriate number (i.e., $M$) of square subarrays in subsequent numerical simulations.

In the numerical results presented below, for single-MS scenarios, the AoA estimation module and the information fusion module are executed sequentially only once, as iterations to mitigate interference are unnecessary. In multi-MS scenarios, based on our experimental observations, the APPLE algorithm typically converges after several iterations between the two modules. 
In experiments involving multiple MSs, the number of iterations is fixed at five.
}
\subsection{PAE Performance}

For comparison, the exploited benchmark schemes are listed as follows:
\begin{itemize}
    \item \textbf{SOMP+PSO}: A two-stage PAE method. It first utilizes the CS-based simultaneous orthogonal matching pursuit (SOMP) algorithm to estimate the AoAs of the activated MS antennas based on the far-field signal model for the entire BS array \cite{tropp2006algorithms}. Then, based on the topological knowledge of the activated MS antennas, the particle swarm optimization (PSO) algorithm is used to search for the positions of the antennas \cite{kennedy1995particle}. Finally, the MS's position and attitude are estimated using the least squares method and the singular value decomposition (SVD) based method, respectively \cite{zhouDoABasedRigidBody2019}.
    \item {\color{black}\textbf{MUSIC+PSO}: A two-stage PAE method similar to ``SOMP+PSO'', with the AoA estimation step replacing the ``SOMP'' algorithm with the classical multiple signal classification (MUSIC) algorithm \cite{schmidtMultipleEmitterLocation1986}.}
    \item \textbf{SOMP+SDR}: A two-stage PAE method similar to ``SOMP+PSO''. After obtaining the AoAs of the activated MS antennas using the SOMP algorithm, a method based on positive semidefinite relaxation (SDR) is employed to estimate MS's position and attitude \cite{wangInvestigationSolutionAngle2020}.
    \item \textbf{LB}: The estimation-error lower bound derived in Sec. \ref{SecLB}.
\end{itemize}


\begin{figure}[t]
    \centering
    \includegraphics[width=.98\linewidth]{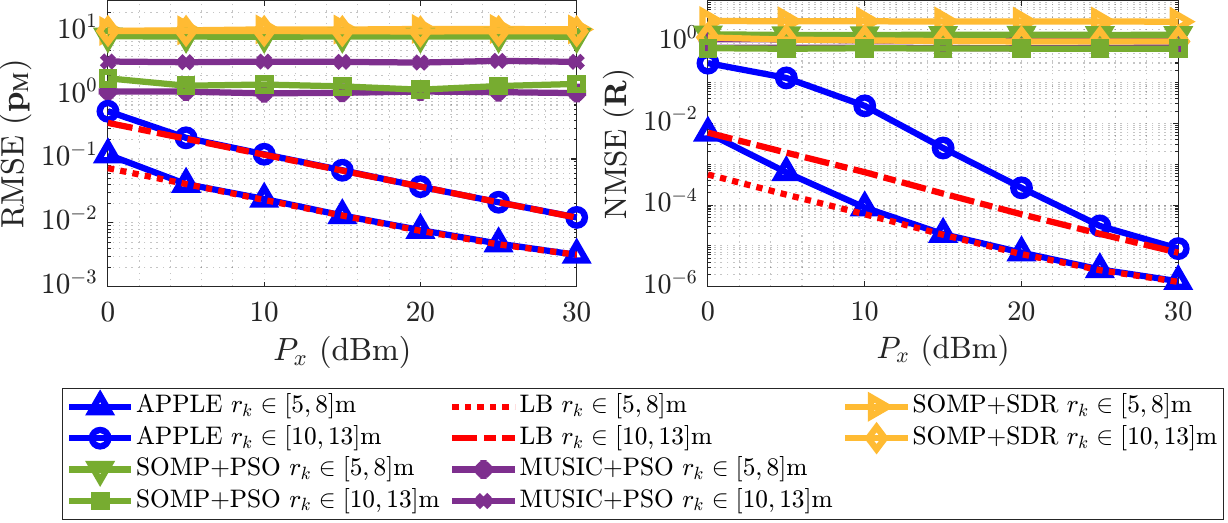}
    \caption{PAE performance with different BS-MS distance, $K=3$.}
    \label{MUVerDistance}
\end{figure}

We first compare the PAE performance by APPLE with those of other baseline methods. In Fig. \ref{MUVerDistance}, we consider $K=3$, and the BS-MS distance $r_k$ are randomly drawn from the ranges $r_k \in [5,8]$m and $r_k \in [10,13]$m. The BS array size is set to $N_{\mathsf{B}}=120$, and the MS array size is set to $N_{\mathsf{M}}=100$. The BS array is uniformly partitioned into $M=16$ subarrays. The left and right subfigures respectively show the estimation performance of MS's position and attitude with the transmit signal power $P_x$. For ``SOMP+PSO'', {\color{black}``MUSIC+PSO''}, and ``SOMP+SDR'', the AoA searching resolution is set to $0.001$rad. The particle swarm size is set to $500$ for ``SOMP+PSO''.
{\color{black}As shown in Fig. \ref{MUVerDistance}, the baseline methods ``SOMP+PSO'', ``MUSIC+PSO'', and ``SOMP+SDR'' struggle to estimate MS's position and attitude in both $r_k \in [5,8]$m and $r_k \in [10,13]$m scenarios (in the attitude estimation subplots, the curves corresponding to ``MUSIC+PSO'' are covered by those of other benchmark methods).} This is because under the near-field signal model, the far-field AoA estimation algorithms SOMP and MUSIC exhibits significant performance degradation due to the model mismatch. Consequently, the PSO and SDR algorithms cannot accurately recover the positions of the activated MS antennas. In contrast, under the two BS-MS distance settings, the position and attitude estimation errors by APPLE decrease smoothly with the increase of $P_x$, and gradually approach the corresponding LBs. For $r_k \in [5,8]$m, the positioning error can be reduced to less than $1$cm with $P_x = 20$dBm.

\begin{figure}[t]
    \centering
    \includegraphics[width=.98\linewidth]{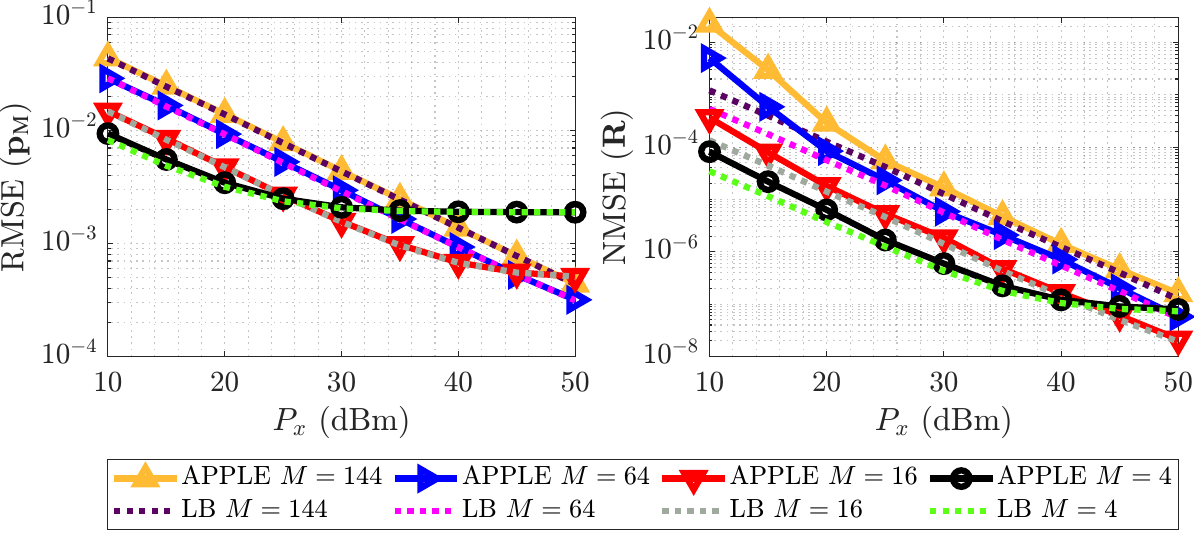}
    \caption{PAE performance by APPLE with a varying number of BS subarrays.}
    \label{verPartNum}
\end{figure}

In Fig. \ref{verPartNum}, we evaluate the performance of the APPLE algorithm under a varying number of BS subarrays, with $M\in \{4,16,64,144\}$, $K=3$, $N_{\mathsf{B}}=120$, and $N_{\mathsf{M}}=60$. The BS-MS distance $r_k$ is randomly drawn from $[5,8]$m. We observe that with a relatively small $P_{x}$, fewer BS subarrays yields a better PAE performance. This is because APPLE not only estimates the position and attitude of the MS but also estimates the complex coefficients $\varrho_{m,(k,t)}$, which is trivial to the considered localization problem. With fewer subarrays, fewer nuisance parameters need to be estimated, leading to improved PAE performance. However, as the $P_x$ increases, APPLE suffers from estimation-error floors with $M=4$ and $M=16$. This is because with a relatively small number of subarrays, the size of each subarray remains large, thereby invalidating the SWFF assumption. Consequently, significant AoA estimation errors occur in the AoA estimation module. Selecting a large number of subarrays ($M=64$ or $M=144$) can mitigate this issue.

\begin{figure}[t]
    \centering
    \includegraphics[width=.98\linewidth]{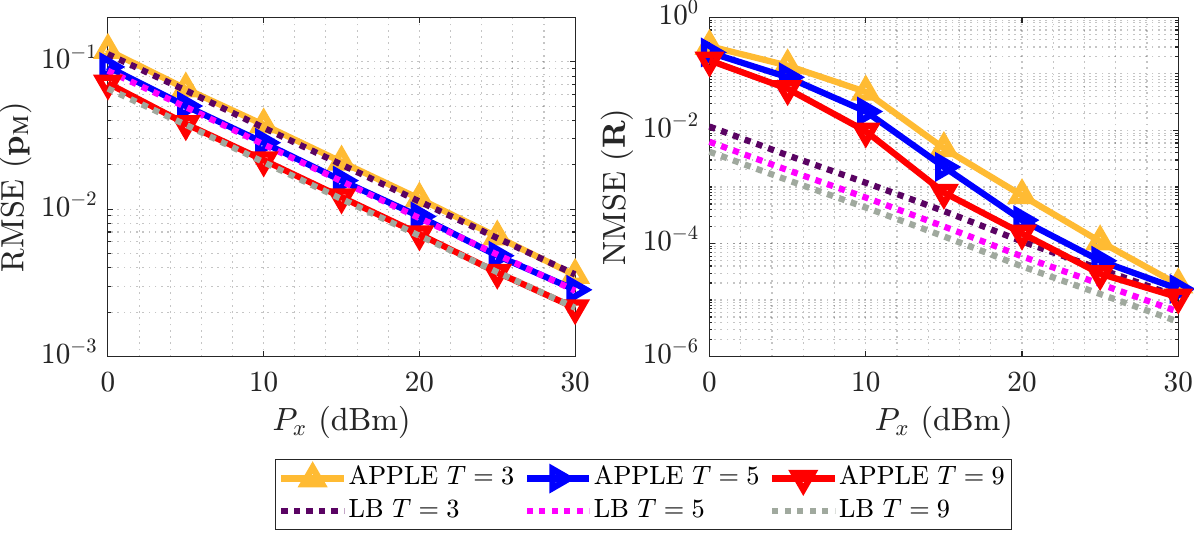}
    \caption{PAE performance by APPLE with a varying number of activated MS antennas.}
    \label{verActAntNum}
\end{figure}

In Fig. \ref{verActAntNum}, we evaluate the performance of the APPLE algorithm under different numbers of activated antennas at each MS, with $K=1$, $N_{\mathsf{B}}=90$, $N_{\mathsf{M}}=40$, $M=9$, and $T\in\{3,5,9\}$. The BS-MS distance $r_k$ is randomly drawn from the range $[5,8]$m. For $T=3$ and $T=9$, the transmit patterns are respectively set to $\mathcal{T}_3 = \left\{(1,1), (1,N_{\mathsf{M}}),(N_{\mathsf{M}},\lceil \frac{N_{\mathsf{M}}-1}{2} \rceil) \right\}$ and $\mathcal{T}_9 = \mathcal{T}_5 \cup \left\{ (1, \lceil \frac{N_{\mathsf{M}}-1}{2} \rceil),(N_{\mathsf{M}},\lceil \frac{N_{\mathsf{M}}-1}{2} \rceil),(\lceil \frac{N_{\mathsf{M}}-1}{2} \rceil,1),( \lceil \frac{N_{\mathsf{M}}-1}{2} \rceil,N_{\mathsf{M}}) \right\}$. This shows that increasing the number of MS antennas activated for signal transmission improves the position and attitude estimation performance of APPLE. When only a small number of antennas are activated (e.g., at $T=3$), APPLE still performs well.

\begin{figure}[t]
    \centering
    \includegraphics[width=.98\linewidth]{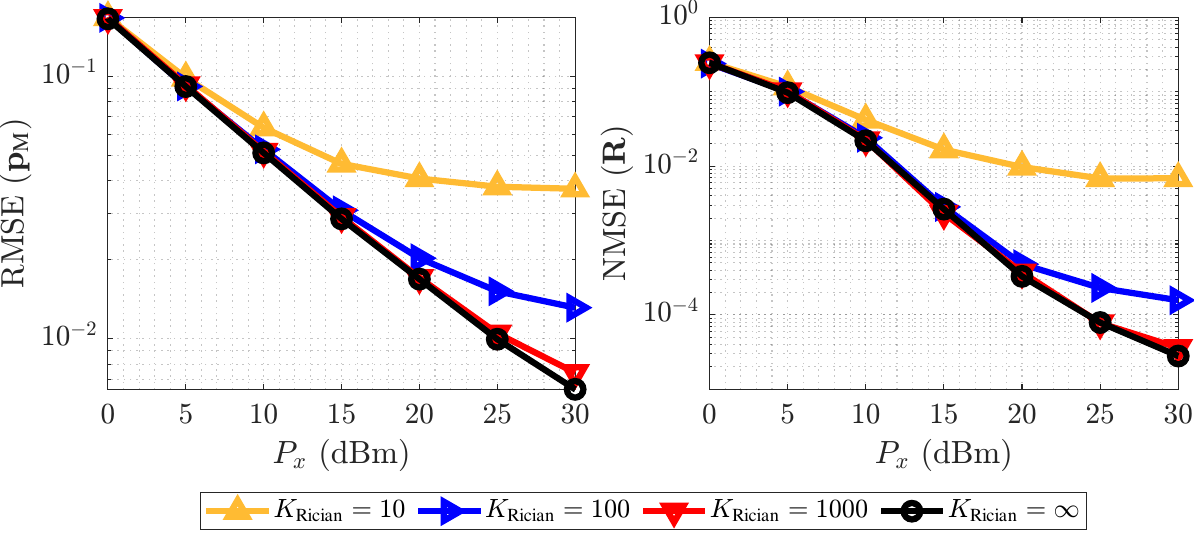}
    \caption{\color{black}PAE performance by APPLE with a varying Rician K-factors.}
    \label{kfac2}
\end{figure}

{\color{black}
In Fig. \ref{kfac2}, we evaluate the impact of NLoS channel component on the APPLE algorithm. We consider the channel between the BS and an activated MS antenna given by \cite{wangWirelessEnergyTransfer2021}
\begin{equation}
    \boldsymbol{h} = \boldsymbol{h}_{\mathrm{LoS}} + \boldsymbol{h}_{\mathrm{NLoS}},
  \end{equation}
where $\boldsymbol{h}_{\mathrm{LoS}}$ and $\boldsymbol{h}_{\mathrm{NLoS}}$ are the LoS and the NLoS components of the channel, respectively. The modelling of $\boldsymbol{h}_{\mathrm{LoS}}$ follows the LoS channel model in Sec. II. B. For the NLoS component, each channel coefficient of $\boldsymbol{h}_{\mathrm{NLoS}}$ is modeled as a CSCG random variable, i.e., $h_{\mathrm{NLoS};q}\sim \mathcal{CN}(0,\frac{\sigma_q^2}{K_{\mathsf{Rician}}})$, where $K_{\mathsf{Rician}}$ is the Rician K-factor, and $\sigma_q^2 = \mathbb{E}\{h_{\mathrm{LoS};q}h_{\mathrm{LoS};q}^*\}$ is the power gain of the LoS channel between the $q$-th BS antenna and the activated MS antenna, $q\in \mathcal{I}_{N_{\mathsf{B}}}$. We evaluate the performance of the APPLE algorithm under different Rician K-factors. The number of MS is set to $K=3$. The transmit pattern is set to the default $\mathcal{T}_5$. Other system settings same as in Fig. \ref{verActAntNum}. The simulations test $K_{\mathsf{Rician}}$ of $10$, $100$, $1000$, and an ideal scenario with no NLoS component (i.e., $K_{\mathsf{Rician}}=\infty$). As shown in Fig. \ref{kfac2}, due to the presence of the NLoS channel component, the APPLE algorithm exhibits an error floor for both position and attitude estimation at high SNR. However, with a large K-factor (typically $K_{\mathsf{Rician}}\geq 100$ for millimeter-wave and terahertz channels \cite{liuLineofSightSpatialModulation2016}), the performance degradation from the NLoS components is less significant.
}

\section{Conclusions}\label{SecConclusion}
In this paper, we tackled a challenging near-field PAE task, where a single BS is employed to estimate the attitudes and positions of multiple MSs based on received signals and the topological information of the activated MS antennas. Given the difficulty of this problem under the original near-field signal model, we presented a SWFF signal model. In this model, the BS array is partitioned into multiple subarrays, with the MS antennas located in the FFR of each BS subarrays. By exploiting the AoAs of the activated MS antennas at different BS subarrays, we formulated the PAE problem under the Bayesian inference framework. We proposed a message passing based algorithm, named APPLE, to solve this PAE problem. Extensive numerical results demonstrated that the APPLE algorithm achieves superior performance in both attitude and position estimation.

{
\color{black}
Looking forward, there are many potential research directions based on the findings of this paper. For instance, the partitioning strategy and the number of subarrays in the BS array, as well as the selection of activated MS antennas (i.e., the design of transmit patterns), merit further exploration. Moreover, IRS has shown great potential for enhancing positioning accuracy by reshaping the wireless channel \cite{dardariNLOSNearFieldLocalization2022b,panRISAidedNearFieldLocalization2023,tengBayesianUserLocalization2022}. Our work can be extended to IRS-assisted localization systems, where variations in AoA across different IRS regions could be exploited for joint positioning and attitude estimation. A key challenge in this extension is to infer the signal's AoA at the IRS indirectly from the BS. Moreover, since APPLE is based solely on AoA, integrating additional measurements from the MS, such as ToA or angular velocity, can further improve positioning and attitude estimation. We leave these interesting research topics as our future work.

}

\appendices

\ifCLASSOPTIONcaptionsoff
  \newpage
\fi

\bibliographystyle{IEEEtran}
\bibliography{IEEEexample.bib}

\begin{thebibliography}{10}
\providecommand{\url}[1]{#1}
\csname url@samestyle\endcsname
\providecommand{\newblock}{\relax}
\providecommand{\bibinfo}[2]{#2}
\providecommand{\BIBentrySTDinterwordspacing}{\spaceskip=0pt\relax}
\providecommand{\BIBentryALTinterwordstretchfactor}{4}
\providecommand{\BIBentryALTinterwordspacing}{\spaceskip=\fontdimen2\font plus
\BIBentryALTinterwordstretchfactor\fontdimen3\font minus
  \fontdimen4\font\relax}
\providecommand{\BIBforeignlanguage}[2]{{%
\expandafter\ifx\csname l@#1\endcsname\relax
\typeout{** WARNING: IEEEtran.bst: No hyphenation pattern has been}%
\typeout{** loaded for the language `#1'. Using the pattern for}%
\typeout{** the default language instead.}%
\else
\language=\csname l@#1\endcsname
\fi
#2}}
\providecommand{\BIBdecl}{\relax}
\BIBdecl

\bibitem{garciaDirectLocalizationMassive2017}
N.~Garcia \emph{et~al.}, ``Direct localization for massive {MIMO},'' \emph{IEEE
  Trans. Signal Process.}, vol.~65, no.~10, pp. 2475--2487, May 2017.

\bibitem{savicFingerprintingBasedPositioningDistributed2015}
V.~Savic \emph{et~al.}, ``Fingerprinting-based positioning in distributed
  massive {MIMO} systems,'' in \emph{2015 IEEE 82nd Vehicular Technology
  Conference (VTC2015-Fall)}, Sep. 2015, pp. 1--5.

\bibitem{tengBayesianUserLocalization2022}
B.~Teng \emph{et~al.}, ``Bayesian user localization and tracking for
  reconfigurable intelligent surface aided {MIMO} systems,'' \emph{IEEE J. Sel.
  Topics Signal Process.}, vol.~16, no.~5, pp. 1040--1054, 2022.

\bibitem{heLargeIntelligentSurface2020a}
J.~He \emph{et~al.}, ``Large intelligent surface for positioning in millimeter
  wave {MIMO} systems,'' in \emph{2020 IEEE 91st Vehicular Technology
  Conference (VTC2020-Spring)}, May 2020, pp. 1--5.

\bibitem{rinchiCompressiveNearFieldLocalization2022a}
O.~Rinchi \emph{et~al.}, ``Compressive near-field localization for multipath
  {RIS}-aided environments,'' \emph{IEEE Commun. Lett.}, vol.~26, no.~6, pp.
  1268--1272, Jun. 2022.

\bibitem{dardariNLOSNearFieldLocalization2022b}
D.~Dardari \emph{et~al.}, ``{LOS/NLOS} near-field localization with a large
  reconfigurable intelligent surface,'' \emph{IEEE Trans. Wireless Commun.},
  vol.~21, no.~6, pp. 4282--4294, Jun. 2022.

\bibitem{wangNearFieldIntegratedSensing2023}
Z.~Wang \emph{et~al.}, ``Near-field integrated sensing and communications,''
  \emph{IEEE Commun. Lett.}, vol.~27, no.~8, pp. 2048--2052, Aug. 2023.

\bibitem{huaNearField3DLocalization2024}
H.~Hua \emph{et~al.}, ``Near-field {3D} localization via {MIMO} radar:
  Cram{\'e}r-rao bound analysis and estimator design,'' \emph{IEEE Trans.
  Signal Process.}, vol.~72, pp. 3879--3895, 2024.

\bibitem{elzanatyReconfigurableIntelligentSurfaces2021a}
A.~Elzanaty \emph{et~al.}, ``Reconfigurable intelligent surfaces for
  localization: Position and orientation error bounds,'' \emph{IEEE Trans.
  Signal Process.}, vol.~69, pp. 5386--5402, 2021.

\bibitem{friedlanderLocalizationSignalsNearField2019}
B.~Friedlander, ``Localization of signals in the near-field of an antenna
  array,'' \emph{IEEE Trans. Signal Process.}, vol.~67, no.~15, pp. 3885--3893,
  Aug. 2019.

\bibitem{panRISAidedNearFieldLocalization2023}
Y.~Pan \emph{et~al.}, ``{RIS}-aided near-field localization and channel
  estimation for the terahertz system,'' \emph{IEEE Journal of Selected Topics
  in Signal Processing}, vol.~17, no.~4, pp. 878--892, Jul. 2023.

\bibitem{selvanFraunhoferFresnelDistances2017}
K.~T. Selvan \emph{et~al.}, ``Fraunhofer and {Fresnel} distances: Unified
  derivation for aperture antennas,'' \emph{IEEE Antennas Propag. Mag.},
  vol.~59, no.~4, pp. 12--15, Aug. 2017.

\bibitem{wangTutorialExtremelyLargeScale2024}
Z.~Wang \emph{et~al.}, ``A tutorial on extremely large-scale {MIMO} for {6G}:
  Fundamentals, signal processing, and applications,'' \emph{IEEE Commun. Surv.
  Tutorials}, early access.

\bibitem{saadVision6GWireless2020a}
W.~Saad \emph{et~al.}, ``A vision of {6G} wireless systems: Applications,
  trends, technologies, and open research problems,'' \emph{IEEE Network},
  vol.~34, no.~3, pp. 134--142, May 2020.

\bibitem{cuiNearFieldMIMOCommunications2023}
M.~Cui \emph{et~al.}, ``Near-field {MIMO} communications for {6G}:
  Fundamentals, challenges, potentials, and future directions,'' \emph{IEEE
  Commun. Mag.}, vol.~61, no.~1, pp. 40--46, Jan. 2023.

\bibitem{liangPassiveLocalizationMixed2010a}
J.~Liang \emph{et~al.}, ``Passive localization of mixed near-field and
  far-field sources using two-stage {MUSIC} algorithm,'' \emph{IEEE Trans.
  Signal Process.}, vol.~58, no.~1, pp. 108--120, 2010.

\bibitem{wangConvexRelaxationMethods2019}
G.~Wang \emph{et~al.}, ``Convex relaxation methods for unified near-field and
  far-field {TDOA}-based localization,'' \emph{IEEE Trans. Wireless Commun.},
  vol.~18, no.~4, pp. 2346--2360, Apr. 2019.

\bibitem{sunSolutionAnalysisTDOA2019}
Y.~Sun \emph{et~al.}, ``Solution and analysis of {TDOA} localization of a near
  or distant source in closed form,'' \emph{IEEE Trans. Signal Process.},
  vol.~67, no.~2, pp. 320--335, Jan. 2019.

\bibitem{zhengScalableNearFieldLocalization2023}
Y.~Zheng \emph{et~al.}, ``Scalable near-field localization based on array
  partitioning and angle-of-arrival fusion,'' \emph{arXiv:2312.12342}, Dec.
  2023.

\bibitem{shahmansooriPositionOrientationEstimation2018}
A.~Shahmansoori \emph{et~al.}, ``Position and orientation estimation through
  millimeter-wave {MIMO} in {5G} systems,'' \emph{IEEE Trans. Wireless
  Commun.}, vol.~17, no.~3, 2018.

\bibitem{marchandPoseEstimationAugmented2016a}
E.~Marchand \emph{et~al.}, ``Pose estimation for augmented reality: A hands-on
  survey,'' \emph{IEEE Trans. Vis. Comput. Graphics}, vol.~22, no.~12, pp.
  2633--2651, Dec. 2016.

\bibitem{wangInvestigationSolutionAngle2020}
Y.~Wang \emph{et~al.}, ``An investigation and solution of angle based rigid
  body localization,'' \emph{IEEE Trans. Signal Process.}, vol.~68, pp.
  5457--5472, 2020.

\bibitem{bjornsonMassiveMIMOReality2019a}
E.~Bj{\"o}rnson \emph{et~al.}, ``Massive {MIMO} is a reality{\textemdash}what
  is next?'' \emph{Digital Signal Processing}, vol.~94, pp. 3--20, Nov. 2019.

\bibitem{shahmansooriTrackingPositionOrientation2019}
A.~Shahmansoori \emph{et~al.}, ``Tracking position and orientation through
  millimeter wave lens {MIMO} in {5G} systems,'' \emph{IEEE Signal Process.
  Lett.}, vol.~26, no.~8, pp. 1222--1226, Aug. 2019.

\bibitem{talvitieHighAccuracyJointPosition2019}
J.~Talvitie \emph{et~al.}, ``High-accuracy joint position and orientation
  estimation in sparse {5G} mmwave channel,'' in \emph{2019 IEEE International
  Conference on Communications (ICC)}, Shanghai, China, May 2019, pp. 1--7.

\bibitem{liJointLocalizationOrientation2022a}
J.~Li \emph{et~al.}, ``Joint localization and orientation estimation in
  millimeter-wave {MIMO} {OFDM} systems via atomic norm minimization,''
  \emph{IEEE Trans. Signal Process.}, vol.~70, pp. 4252--4264, 2022.

\bibitem{mendrzikHarnessingNLOSComponents2019a}
R.~Mendrzik \emph{et~al.}, ``Harnessing {NLOS} components for position and
  orientation estimation in {5G} millimeter wave {MIMO},'' \emph{IEEE Trans.
  Wireless Commun.}, vol.~18, no.~1, pp. 93--107, Jan. 2019.

\bibitem{zhouDoABasedRigidBody2019}
B.~Zhou \emph{et~al.}, ``{DoA}-based rigid body localization adopting single
  base station,'' \emph{IEEE Commun. Lett.}, vol.~23, no.~3, pp. 494--497, Mar.
  2019.

\bibitem{wangBiasReducedSemidefinite2021}
G.~Wang \emph{et~al.}, ``Bias reduced semidefinite relaxation method for {3-D}
  rigid body localization using {AOA},'' \emph{IEEE Trans. Signal Process.},
  vol.~69, pp. 3415--3430, 2021.

\bibitem{esfahaniOriNetRobust3D2020}
M.~A. Esfahani \emph{et~al.}, ``Orinet: Robust 3-{D} orientation estimation
  with a single particular {IMU},'' \emph{IEEE Robot. Autom. Lett.}, vol.~5,
  no.~2, pp. 399--406, Apr. 2020.

\bibitem{zhangRobustMethodMeasuring2021}
J.~Zhang \emph{et~al.}, ``Robust method for measuring the position and
  orientation of drogue based on stereo vision,'' \emph{IEEE Trans. Ind.
  Electron.}, vol.~68, no.~5, pp. 4298--4308, May 2021.

\bibitem{limaTrajectoryTrackingControl2016}
T.~A. Lima \emph{et~al.}, ``Trajectory tracking control of a mobile robot using
  lidar sensor for position and orientation estimation,'' in \emph{2016 12th
  IEEE International Conference on Industry Applications}, Nov. 2016, pp. 1--6.

\bibitem{zhangMicromagnetometerCalibrationAccurate2015}
Z.-Q. Zhang \emph{et~al.}, ``Micromagnetometer calibration for accurate
  orientation estimation,'' \emph{IEEE Trans. Biomed. Eng.}, vol.~62, no.~2,
  pp. 553--560, Feb. 2015.

\bibitem{kokUsingInertialSensors2018}
M.~Kok \emph{et~al.}, ``Using inertial sensors for position and orientation
  estimation,'' \emph{arXiv:2310.01342}, 2017.

\bibitem{meyerMessagePassingAlgorithms2018}
F.~Meyer \emph{et~al.}, ``Message passing algorithms for scalable multitarget
  tracking,'' \emph{Proc. IEEE}, vol. 106, no.~2, pp. 221--259, Feb. 2018.

\bibitem{meyerDistributedLocalizationTracking2016}
------, ``Distributed localization and tracking of mobile networks including
  noncooperative objects,'' \emph{IEEE Trans. Signal Inf. Process. Netw.},
  vol.~2, no.~1, pp. 57--71, Mar. 2016.

\bibitem{yuanCooperativeJointLocalization2016}
W.~Yuan \emph{et~al.}, ``Cooperative joint localization and clock
  synchronization based on {Gaussian} message passing in asynchronous wireless
  networks,'' \emph{IEEE Transactions on Vehicular Technology}, vol.~65, no.~9,
  pp. 7258--7273, Sep. 2016.

\bibitem{liuLineofSightSpatialModulation2016}
P.~Liu \emph{et~al.}, ``Line-of-sight spatial modulation for indoor {mmWave}
  communication at 60 {GHz},'' \emph{IEEE Trans. Wireless Commun.}, vol.~15,
  no.~11, pp. 7373--7389, Nov. 2016.

\bibitem{luTutorialNearFieldXLMIMO2024}
H.~Lu \emph{et~al.}, ``A tutorial on near-field {XL-MIMO} communications
  towards {6G},'' \emph{IEEE Commun. Surv. Tutorials}, pp. 1--1, 2024.

\bibitem{bohagenDesignOptimalHighRank2007a}
F.~Bohagen \emph{et~al.}, ``Design of optimal high-rank line-of-sight {MIMO}
  channels,'' \emph{IEEE Trans. Wireless Commun.}, vol.~6, no.~4, pp.
  1420--1425, Apr. 2007.

\bibitem{jammalamadaka2001topics}
S.~Jammalamadaka, \emph{Topics in Circular Statistics}.\hskip 1em plus 0.5em
  minus 0.4em\relax World Scientific, 2001, vol. 336.

\bibitem{badiuVariationalBayesianInference2017}
M.-A. Badiu \emph{et~al.}, ``Variational {Bayesian} inference of line
  spectra,'' \emph{IEEE Trans. Signal Process.}, vol.~65, no.~9, pp.
  2247--2261, May 2017.

\bibitem{kschischangFactorGraphsSumproduct2001}
F.~Kschischang \emph{et~al.}, ``Factor graphs and the sum-product algorithm,''
  \emph{IEEE Trans. Inf. Theory}, vol.~47, no.~2, pp. 498--519, Feb. 2001.

\bibitem{bishop2006pattern}
C.~M. Bishop \emph{et~al.}, \emph{Pattern Recognition and Machine
  Learning}.\hskip 1em plus 0.5em minus 0.4em\relax Springer, 2006, vol.~4,
  no.~4.

\bibitem{petersenMatrixCookbook}
K.~B. Petersen \emph{et~al.}, ``The matrix cookbook,'' \emph{Technical
  University of Denmark}, vol.~7, no.~15, p. 510, 2008.

\bibitem{fortunatiPerformanceBoundsParameter2017}
S.~Fortunati \emph{et~al.}, ``Performance bounds for parameter estimation under
  misspecified models: Fundamental findings and applications,'' \emph{IEEE
  Signal Process. Mag.}, vol.~34, no.~6, pp. 142--157, Nov. 2017.

\bibitem{chenChannelModelMismatch2022}
H.~Chen \emph{et~al.}, ``Channel model mismatch analysis for {XL-MIMO} systems
  from a localization perspective,'' in \emph{2022 IEEE Global Communications
  Conference}, Rio de Janeiro, Brazil, Dec. 2022, pp. 1588--1593.

\bibitem{TR38901}
``Study on channel model for frequencies from 0.5 to 100 {GHz},'' \emph{3GPP TR
  38.901 version 16.1.0 Release 16}, 2020.

\bibitem{tropp2006algorithms}
J.~A. Tropp \emph{et~al.}, ``Algorithms for simultaneous sparse approximation.
  part {I}: Greedy pursuit,'' \emph{Signal processing}, vol.~86, no.~3, pp.
  572--588, 2006.

\bibitem{kennedy1995particle}
J.~Kennedy \emph{et~al.}, ``Particle swarm optimization,'' in \emph{Proceedings
  of ICNN'95-international conference on neural networks}, vol.~4.\hskip 1em
  plus 0.5em minus 0.4em\relax IEEE, 1995, pp. 1942--1948.

\bibitem{schmidtMultipleEmitterLocation1986}
R.~Schmidt, ``Multiple emitter location and signal parameter estimation,''
  \emph{IEEE Trans. Antennas Propag.}, vol.~34, no.~3, pp. 276--280, 1986.

\bibitem{wangWirelessEnergyTransfer2021}
J.~Wang \emph{et~al.}, ``Wireless energy transfer in extra-large massive {MIMO}
  {Rician} channels,'' \emph{IEEE Trans. Wireless Commun.}, vol.~20, no.~9, pp.
  5628--5641, Sep. 2021.

\bibitem{zhuGridlessVariationalBayesian2019}
J.~Zhu \emph{et~al.}, ``Grid-less variational {Bayesian} line spectral
  estimation with multiple measurement vectors,'' \emph{Signal Processing},
  vol. 161, pp. 155--164, Aug. 2019.

\end{thebibliography}





\end{document}